\documentstyle[aas2pp4,psfig,overrulehere]{article}  % two-column style
\psdraft
 \def\dps{\displaystyle} \def\scs{\scriptstyle}
\def\scscs{\scriptscriptstyle}
\def\fracdps#1#2{\frac{\dps #1}{\dps #2}}

\def\vec#1{{\bf #1}}

\def\defdef{\;{\buildrel \rm def \over =}\;}
\def\etal{{et al.}}
\def\Sec{\S} % \def\Sec{{Sec.~}}   

\def\deg{{\rm\,o}}

\def\Bvec{{\vec{B}}}

\def\Btranvec{{\vec{B}_t^{}}}
\def\Blong{{B_l}}   
\def\Blongmax{{\Blong}_0^{}} 
\def\Btran{{B_t}}   \def\btran{\Btran}

\def\Bbl#1{{B^{#1}_{bl}}}
\def\betabl{{\beta_{bl}^{}}}

\def\FLtranvec{{{\vec{F}_{L{\scscs t}}}}}

\def\Jvec{{\vec{J}}} 
\def\Jlong{{J_l^{}}}
\def\Jlongvec{{\vec{J}_l^{}}}
\def\jlong{\Jlong}
\def\Jlongvec{{\vec{J}_l^{}}}

\def\ephi{{\vec{e}_{\phi}}}

\def\vrel#1{{v_{rel}^{#1}}}
\def\vrelt#1{{v_{rel}^{#1}}}
\def\vrelvec{{\vec{v}_{rel}^{}}}
\def\vreltvec{{\vec{v}_{rel}^{}}}
\def\vterm#1{{v^{#1}_{\scs term}}}
\def\vapex#1{{v^{#1}_{\scs apex}}}
\def\vapexvec{{\vec{v}^{}_{\scs apex}}}
\def\vrise#1{{v^{#1}_{\scs rise}}}

\def\valfven{{v_{A}^{}}}
\def\vsound#1{{c_s^{#1}}}

\def\vortlong{{\omega_l}}
\def\vortlongvec{{\vec{\omega}_l}}
\def\vortl{\vortlong}

\def\pext{{p_e^{}}}
\def\pextf{{p_{ef}^{}}}
\def\rhoext{{\rho_e^{}}}
\def\Text{{T_e^{}}}
\def\hp{{H_p^{}}} 

\def\dpre{{\Delta p}}
\def\drho{{\Delta\rho}}

\def\ekin{{e_{\scs kin}^{}}}
\def\ekinas{{e_{\scs k}^{}}_{as}}
\def\etran{{e_{{\scs mag}_{\scscs t}^{}}^{}}}
\def\rtube{{R}}
\def\tilder{{\tilde{\rtube}}}
\def\Lbl{{L_{bl}^{}}}
\def\Lwake{{L_{wake}^{}}}
\def\csarea{{S}} %A_{cs}^{}}}
\def\csdiam{{d}}
\def\Max{{mt}} %\def\Max{{M}}
\def\cp{{c_p}}
\def\CD{{C_D^{}}}
\def\Psimax{{\Psi_{\Max}}}

\def\Rm#1{{Re_m^{#1}}}  \def\Rem#1{\Rm{#1}}

\begin{document}
%%%%%%%%%%%%%%%%%%%%%%%%%%%%%%%%%%%%%%%%%%%%%%%%%%%%%%%%%%%%%%%%%%%%%%
%%%%%%%%%%%%%%%%%%%%%%%%%%%%%%%%%%%%%%%%%%%%%%%%%%%%%%%%%%%%%%%%%%%%%%
%
\def\rahmen#1#2{\vbox{\hrule\hbox{\vrule\hskip#1\vbox{\vskip#1\relax
\hbox to \hsize{\hss #2 \hss} \vskip #1}\hskip#1\vrule}\hrule}}

\vspace{-8.5cm}\ \hskip 0cm
\centerline{\vbox to 0pt{\hsize 12.0cm $$\rahmen{0.2cm}{\noindent 
To be published in  
{\it The Astrophysical Journal}, Vol.~{\bf 492}, Jan 10th, 1998}
$$\vss}}
\vskip 2cm
%
%%%%%%%%%%%%%%%%%%%%%%%%%%%%%%%%%%%%%%%%%%%%%%%%%%%%%%%%%%%%%%%%%%%%%%
%%%%%%%%%%%%%%%%%%%%%%%%%%%%%%%%%%%%%%%%%%%%%%%%%%%%%%%%%%%%%%%%%%%%%%

\title{The physics of twisted magnetic tubes rising in a stratified
medium: two dimensional results}

\author{Emonet, T., Moreno-Insertis, F.}
\affil{Instituto de Astrof\'\i sica de Canarias, 38200 La Laguna (Tenerife),
Spain} 
\affil{temonet@ll.iac.es, fmi@ll.iac.es}

%%%%%%%%%%%%%%%%%%%%%%%%%%%%%%%%%%%%%%%%%%%%%%%%%%%%%%%%%%%%%%%%%%%%%%
%%%%%%%%%%%%%%%%%%%%%%%%%%%%%%%%%%%%%%%%%%%%%%%%%%%%%%%%%%%%%%%%%%%%%%

\begin{abstract}
The physics of a twisted magnetic flux tube rising in a stratified medium is
studied using a numerical MHD code. The problem considered is fully
compressible (no Boussinesq approximation), includes ohmic resistivity, and
is two dimensional, i.e., there is no variation of the variables in the
direction of the tube axis. We study a high plasma $\beta$ case with small
ratio of radius to external pressure scaleheight. The results obtained can
therefore be of relevance to understand the transport of magnetic flux across
the solar convection zone.

We confirm that a sufficient twist of the field lines around the tube axis
can suppress the conversion of the tube into two vortex rolls. For a tube
with relative density deficit of order $1/\beta$ (the classical Parker
buoyancy) and radius smaller than a pressure scale-height ($\rtube^2 \ll
\hp^2$), the minimum amount of twist necessary corresponds to an average
pitch angle of order $\sin^{-1}[(\rtube/\hp)^{1/2}]$. The evolution of a tube
with this degree of twist is studied in detail, including the initial
transient phase, the internal torsional oscillations and the asymptotic,
quasi-stationary phase. During the initial phase, the outermost, weakly
magnetized layers of the tube are torn off its main body and endowed with
vorticity. They yield a trailing magnetized wake with two vortex rolls. Which
fraction of the total magnetic flux is brought to the wake is a function of
the initial degree of twist. In the weakly twisted case, most of the initial
tube is turned into vortex rolls. At the opposite end (strong initial twist),
the tube rises with only a small deformation and no substantial loss of
magnetic flux. The formation of the wake and the loss of flux from the main
body of the tube are basically complete after the initial transient phase.

A sharp interface between the tube interior and the external flows is formed
at the tube front and sides; it has the characteristic features of a magnetic
boundary layer. Its structure is determined as an equilibrium between ohmic
diffusion and field advection through the external flows. It is the site of
vorticity generation via the magnetic field during the whole tube evolution.

From the hydrodynamical point of view, this problem constitutes an
intermediate case between the rise of air bubbles in water and the motion of
a rigid cylinder in an external medium. As in the first one, the tube is
deformable and the outcome of the experiment (shape of rising object and
wake) depends on the value of the Weber number. Several structural features
obtained in the present simulation are also observed in rising air bubbles,
like a central {\it tail}, and a {\it skirt} enveloping the wake. As in rigid
cylinders, the boundary layer satisfies a no-slip condition (provided for in
the tube by the magnetic field), and secondary rolls are formed at the
lateral edges of the moving object.
\end{abstract}

%%%%%%%%%%%%%%%%%%%%%%%%%%%%%%%%%%%%%%%%%%%%%%%%%%%%%%%%%%%%%%%%%%%%%%
%%%%%%%%%%%%%%%%%%%%%%%%%%%%%%%%%%%%%%%%%%%%%%%%%%%%%%%%%%%%%%%%%%%%%%

%%%%%%%%%%%%%%%%%%%%%%%%%%%%%%%%%%%%%%%%%%%%%%%%%%%%%%%%%%%%%%%%%%%%%%
%%%%%%%%%%%%%%%%%%%%%%%%%%%%%%%%%%%%%%%%%%%%%%%%%%%%%%%%%%%%%%%%%%%%%%
\section{Introduction}
\label{sec_intro}

The rise of magnetic flux from the deep levels of the solar convection
zone to the photosphere is a complex phenomenon involving many different
magnetic and hydrodynamical processes. Particular attention has been devoted
in the past $20$ years to the time-evolution of a single buoyant magnetic
flux tube considered as a one-dimensional object (\cite{morenoinsertis86};
\cite{choudhuri89}; \cite{dsilvachoudhuri93}; \cite{fanetal93};
\cite{fanetal94}; \cite{caligarietal95}. Further references and a recent
review can be found in the paper by \cite{morenoinsertis97a}). These
calculations incorporate several aspects of the basic physics of the rise of
the magnetic tubes (buoyancy, magnetic and rotational forces, external
stratification, etc). They have been successful in predicting morphological
and kinematic features of the resulting active regions which are observed at
the surface of the Sun.

Yet, the assumption of one-dimensionality of the magnetic region is certainly
a drastic simplification. From laboratory and numerical experiments in
different contexts we know of the complicated hydrodynamical and magnetic
structure within and outside a tubular object which is moving with respect to
the surrounding fluid. The one-dimensional numerical models mentioned above,
in particular, do not contain two ingredients which turn out to be
fundamental in studying the rise of buoyant magnetized plasma regions, to
wit, the vorticity of the velocity field and the twist of the field lines
around the main axis of the tube.  

Vorticity and transverse field components may be crucial for the formation of
the tubes in the first place (\cite{cattaneohughes88}; \cite{cattaneoetal90};
\cite{matthewsetal95}). They also play a central role in the time evolution
of the rising magnetic region. A clear warning in this sense came from the
work of Sch\"ussler (1979), who showed how the cross section of a straight,
buoyant magnetic tube initially with the same temperature of its surroundings
develops an {\it umbrella} shape (two side lobes connected on their upper
side by an arch). The side lobes rotate in opposite directions around a
horizontal axis, each thus constituting a vortex tube; they finally detach
from each other and from the arch above them.  The whole process occurs at
the beginning of the rise, namely before the tube has risen across a height
equivalent to a few times its own diameter. The physics involved has been
considered in detail by Longcope, Fisher \& Arendt (1996). These authors have
studied the Boussinesq problem, including untwisted and very weakly twisted
magnetic tubes.  They clearly show how the two rotating side lobes, when
detached from the rest, are subjected to a downward-pointing lift force, as a
result of their flow being non-circulation free. The lift ends up cancelling
the buoyancy force, this being the reason for their horizontal asymptotic
motion. If this were a universal mechanism operating on all rising magnetic
flux tubes, then magnetic buoyancy should no longer be considered an
efficient mechanism to bring magnetic flux to the photosphere.

In the present paper we consider in detail the more general case of a buoyant
magnetic flux tube with an arbitrary initial twist of the field lines (but
still horizontal and with uniform values of all variables along the direction
of the axis). The transverse magnetic field (i.e., the component of the field
vector normal to the tube axis) imparts a certain rigidity to the tube cross
section. If strong enough, it can prevent the conversion of the tube into a
vortex tube pair. The minimum amount of twist necessary for that corresponds
to an average pitch angle of order $\sin^{-1}\{[\rtube/\hp\;\beta/2\;
|\Delta\rho/\rho\,|\,]^{1/2}\}$, with $\rtube$ the tube radius, $\hp$ the
external pressure scaleheight, and $\Delta \rho$ the density difference between
tube and surroundings (see \Sec\ref{sec_tube_unity}). This approximate
criterion is indeed fulfilled by magnetic tubes with the classical Parker
magnetic buoyancy, as shown in a preliminary presentation of this paper
(\cite{fmiemo96}). In the present article, we explore in more detail the
physics involved in that process, discussing a number of (M)HD processes
occurring inside the tube, in the boundary layer at its periphery and in the
trailing wake. We also show how the results of Sch\"ussler (1979) and
Longcope \etal (1996) can be seen as the limiting case in which the trailing
wake in fact engulfs most of the original rising tube. The pitch angle just
mentioned thus signals the borderline between the weak and strong twist
regimes: a buoyant tube with an initial twist above that level rises without
being strongly deformed and is followed by a wake containing only a small
fraction of the initial total magnetic flux.

Additionally to the work of Moreno-Insertis \& Emonet (1996), there is
another paper in the recent literature dealing with a related subject
(Cargill et al. 1996). The authors have studied the interaction of a twisted
tube with a magnetized medium in the absence of gravity when the tube is
subjected to an ad-hoc, spatially uniform acceleration. Special emphasis was
put on the reconnection of the ambient magnetic field with the tube's own
one. Buoyancy, stratification, or different degrees of twist were not studied
in that paper. Two further papers, submitted simultaneously with the present
one, deal with the rise of buoyant twisted magnetic tubes (\cite{fanetal97};
Hughes, Falle \& Joarder, 1998). The first 
authors, in particular, study the interaction between tubes rising in pairs.
The results of both papers concerning the rise of single tubes are in general
agreement with those of Moreno-Insertis \& Emonet (1996). 

The layout of the paper is as follows. After a brief presentation of the
equations and the numerical procedure in \Sec\ref{sec_equations}, the basic
features of the physical problem are considered (\Sec\ref{sec_parameters}).
This includes the main parameters and a discussion of the amount of twist
necessary to prevent the deformation of the tube and its conversion into
vortex rolls. In \Sec\ref{sec_rise}, the simulation of the rise of a tube
with that amount of twist is presented. In particular, the initial
acceleration phase, the internal torsional oscillations and the later
asymptotic phase are discussed. Section \Sec\ref{sec_interface} deals with
the structure of the magnetic boundary layer around the tube, and
\Sec\ref{sec_wake} examines the trailing wake.  Finally, the transition
between the twisted and untwisted case is explained in \Sec\ref{sec_split}. A
general discussion follows in \Sec\ref{sec_discussion}.

%%%%%%%%%%%%%%%%%%%%%%%%%%%%%%%%%%%%%%%%%%%%%%%%%%%%%%%%%%%%%%%%%%%%%%
%%%%%%%%%%%%%%%%%%%%%%%%%%%%%%%%%%%%%%%%%%%%%%%%%%%%%%%%%%%%%%%%%%%%%%
\section{Equations and numerical procedure}
\label{sec_equations} 
%

%%%%%%%%%%%%%%%%%%%%%%
\subsection{Equations}
Our medium is an ideal compressible and stratified gas governed by 
the general equations of the magnetohydrodynamics (MHD) including Ohmic
diffusion:  
\begin{eqnarray}
\fracdps{D\rho}{Dt} &=& -\rho \nabla\cdot\vec{v} \;, 
\label{eq_continuity}
\\
\noalign{\vskip 2mm}
 \rho\fracdps{D\vec{v}}{Dt} &=&
-\nabla p +\fracdps{1}{4\pi} \left(\nabla\times\vec{B}\right)\times\vec{B}
+\rho\, \vec{g} \;,
\label{eq_motion}
\\
\noalign{\vskip 2mm}
  \fracdps{\partial\vec{B}}{\partial t} &=&
\vec{\nabla}\times\left(\vec{v}\times\vec{B}\right) 
+ \eta \Delta\vec{B}  \;, 
\label{eq_induction}
\\
\noalign{\vskip 2mm}
\rho\fracdps{D e}{Dt} &=& -p \nabla\cdot\vec{v} + \frac{\eta}{4\pi}
\left(\nabla\times\vec{B}\right)^2 \;,
\label{eq_energy}
\end{eqnarray}
with $e$ the internal energy per unit mass and $\eta$ the ohmic diffusivity,
which is assumed constant.  All other symbols have their customary
meaning. Cartesian coordinates $(x,y,z)$ are adopted so that the
$z$-direction is anti-parallel to $\vec{g}$. In this paper we consider a
two-dimensional problem: we assume that there are no variations of the
physical variables along the $y$-axis ($\partial/\partial y \equiv 0$),
although $B_y$ and $v_y$ are generally non zero.

Equations (\ref{eq_continuity})--(\ref{eq_energy}) are cast in their
conservative form and solved with a code written by \cite{shibata83}. The
latter is based on a modified Lax-Wendroff scheme (\cite{rubinbur67}) and
stabilized with artificial viscosity as described by Richtmyer \& Morton
(1967). The inclusion of physical resistivity in our equations and the
absence of any shock phenomena in the solutions permit us to minimize the use
of the artificial viscosity. The latter is restricted, in any case, to
regions of very steep gradients only.

This code has been repeatedly tested and used for two-dimensional simulations
of the outbreak of the magnetic field at the surface of the Sun by Shibata
and collaborators (see e.g. \cite{shibata83}; \cite{shibata89};
\cite{kaisigetal90}). In addition, we have successfully run several tests for
our problem, checking for the convergence of the code as well as for the
conservation of mass, energy and magnetic flux in the box.

In this paper, we only consider a case with left-right symmetry about a
vertical plane containing the tube axis. For each half of the tube we use a
numerical grid of 300 points in the horizontal direction and 700 points in
the vertical one (although in the figures presented here, only a small
fraction of the box is shown). In the following, the results are given in
dimensionless form, using as units the background density, pressure
scaleheight and Alfv\'en speed calculated at the center of the tube at time
$t=0$.

%%%%%%%%%%%%%%%%%%%%%%%%%%%%%%%
\subsection{Initial conditions}
\label{sec_inicond}
The initial condition consists of an unperturbed background atmosphere, with
pressure $\pext$ and density $\rhoext$, and, superimposed, a perturbation
associated with a magnetic flux tube. To avoid any confusion due to
pseudo-convective effects, the background atmosphere is adiabatically
stratified. It spans vertically $80\%$ of the pressure scaleheight at the
bottom of the box. The pressure contrast between the top and bottom is
$2.6$. The density contrast is $1.8$.

After inclusion of the magnetic tube, the resulting system satisfies the
following simple condition: $-\vec{\nabla}\dpre + \Jvec/c \times\vec{B}=0$,
with $\dpre\defdef p-\pext$ the pressure excess as compared with the
background stratification, $\Jvec$ the electric current density and $c$ the
speed of light. In the absence of gravity, this would be a perfect
equilibrium condition.  The density profile in the magnetic region is
determined by assuming that the entropy in the tube is constant and equal to
the unperturbed value in the atmosphere. The tube at time $t=0$ is thus
buoyant ($\rho < \rhoext $) and fulfills

\begin{equation}
\frac{\drho}{\rho}\defdef\frac{\rho-\rhoext}{\rho} \cong
-\frac{1}{\gamma\beta}\; 
\label{eq_drho0}
\end{equation}

\noindent to first order in $1/\beta$, with $\gamma$ the specific heat
ratio. This case is intermediate between the two extreme possibilities of
full thermal equilibrium, i.e., $T=\Text$ (which would be a factor $\gamma$
more buoyant) and the case of a tube with $\rho = \rhoext$. The evolution
presented here is qualitatively very similar to the first case, whereas it
deviates in important respects from the second.

Along this paper we will deal with a number of different magnetic field
profiles at time $t=0$. The longitudinal field $\Blong=B_y$ will be taken to
have a gaussian profile,
\begin{equation}
\dps \Blong(t=0) \propto \exp(\strut \dps -r^2/\rtube^2)  \;.
\label{eq_blong}
\end{equation}
The transverse field $\Btran(t=0)$ is chosen to be purely azimuthal; calling
$r$ and $\phi$ the polar coordinates around the tube center, we have
$\Btranvec (t=0) = B_\phi \ephi$. For $B_\phi$, we will choose distributions
such that the pitch angle $\Psi$,
\begin{equation}
\Psi \defdef \hbox{atan}\left(\frac{B_\phi^{}}{\Blong}\right)\;,
\label{eq_pitch_angle}
\end{equation}
adopts an asymptotically flat profile:
\begin{equation}
B_\phi^{}  \propto \Blong \,\fracdps{\strut a r^n}{\strut a r^n +\rtube^n}\;,
\label{eq_pitchassympt}
\end{equation} 
or, alternatively, an exponentially decaying one:
\begin{equation}
B_\phi^{}  \propto \Blong \,\left({\fracdps{1}{\sqrt{n}}}
\fracdps{\strut r}{\strut \rtube} 
\right)^{n}\, \exp\,\left(-\fracdps{r^2}{\rtube^2} + \fracdps{n}{2}\right) \;, 
\label{eq_pitchgauss}
\end{equation} 
The choice of $n$ determines the {\it rigidity} of the tube center. In this
paper we consider the case $n=3$ and $a=0.9$. As we will see, both
profiles (\ref{eq_pitchassympt}) and (\ref{eq_pitchgauss}) yield basically
the same time evolution (\S\ref{sec_shearflow}). As explained in the
discussion, we have also run some tests with flat, top-hat magnetic profiles
(\S\ref{sec_adeq_param_distr}).

%
%%%%%%%%%%%%%%%%%%%%%%%%%%%%%%%%
\subsection{Boundary conditions}
\label{sec_boundcond}
The side and bottom boundaries are closed lids. A closed boundary at the top
of the box must be avoided for two reasons: (1) the sound waves generated by
the tube in the external medium must be able to leave the box and (2) the
rise of the tube should not be unduly braked through the excess pressure at a
closed boundary at the top. An ideal {\it free} boundary should be
transparent for the outgoing disturbances and should not introduce
disturbances through incoming waves. In the present calculations we achieve
this by introducing a {\it fiducial layer} (\cite{nord90}) well above the
upper boundary of the box. At each time step, the pressure in the fiducial
layer is calculated by assuming that each point in it is in static
equilibrium with respect to the point on the boundary of the box lying
directly below it; the densities are then obtained via the constant
background entropy; the velocity is set to zero in the fiducial layer. The
net mass flux across the boundary is not necessarily zero. However, the total
mass of the box only varies by a maximum factor of $10^{-6}$ during each
single run.
%%%%%%%%%%%%%%%%%%%%%%%%%%%%%%%%%%%%%%%%%%%%%%%%%%%%%%%%%%%%%%%%%%%%%%
%%%%%%%%%%%%%%%%%%%%%%%%%%%%%%%%%%%%%%%%%%%%%%%%%%%%%%%%%%%%%%%%%%%%%%

%%%%%%%%%%%%%%%%%%%%%%%%%%%%%%%%%%%%%%%%%%%%%%%%%%%%%%%%%%%%%%%%%%%%%%
%%%%%%%%%%%%%%%%%%%%%%%%%%%%%%%%%%%%%%%%%%%%%%%%%%%%%%%%%%%%%%%%%%%%%%
\section{The physical problem: parameters governing the evolution}
\label{sec_parameters}
The present physical problem is characterized by four basic dimensionless
input parameters: (1) the thickness of the tube in units of the local
pressure scale height, $\tilder \defdef \rtube/\hp$; (2) the plasma beta at
the center of the tube, $\beta_0$; (3) the ratio between the transverse and
the longitudinal component of the field, as measured by, e.g., the pitch
angle at a representative position of the tube, $\Psimax$, to be specified in
the following; (4) the ohmic diffusivity $\eta$ in terms of, e.g., $\valfven
\hp$, with $\valfven$ the Alfv\'en speed: $\tilde{\eta} \defdef
\eta/(\valfven\hp)$.

These parameters determine the properties of the initial magnetic tube and
are independent of the velocity of the flow that develops along
time. However, in this problem there is a characteristic value for the flow
velocity, namely the terminal speed of rise, $\vterm{}$, given by the
dynamical equilibrium between buoyancy and aerodynamic drag:
\begin{equation}
\vterm{2} \sim \frac{\pi}{\gamma}\; \frac{\tilder}{\CD}
\;\frac{\Delta\rho}{\rho}  \; \vsound{2} \;, 
\label{eq_term_speed_order}
\end{equation}
with $C_D$ the customary drag coefficient, $C_D \sim \hbox{O}(1)$, and
$\vsound{}$ the sound speed. Hence, the customary velocity-related
parameters, like, e.g., the Mach number, the Reynolds number and the magnetic
Weber number, can be immediately obtained as a function of the foregoing
input parameters. For instance, using equation~(\ref{eq_drho0}) we obtain
that the Mach number must be of order: $M^2 = \hbox{O}(\tilder/\beta_0)$. As
a result, we expect the rise to be very subsonic and, assuming $\tilder \ll
1$, sub-Alfvenic as well. This latter condition can be violated for tubes
rising to higher levels of the convection zone.

The rest of this section is devoted to a discussion of the values expected
for the most important parameters of this problem.

%%%%%%%%%%%%
\subsection{The amount of twist and the deformation of the 
tube}\label{sec_tube_unity}  

The central parameter for the present paper is the twist of the tube as
measured by the pitch angle of the field lines, $\Psi$. The values of this
parameter of interest for our calculation are those for which the transverse
component of the field, $\Btranvec$, is able to suppress the conversion of
the tube into a pair of vortex rolls. In the following we study different
agents that tend to deform the tube.  We come to the conclusion that the
minimum value of $\Btran$ necessary to counteract them is the same for all.

\subsubsection{Differential buoyancy and pressure fluctuations along the
boundary}\label{sec_diff_buoy_pres_fluct}

The tube of the initial condition explained in \Sec\ref{sec_inicond} is
increasingly buoyant toward its center. Thus, the central regions rise faster
than the periphery; the upper layers of the tube are thereby compressed
whereas those located below the center are expanded. The magnetic field lines
are deformed by this process; the transverse field, in particular, exerts an
increasing resistance against further deformation.  One can calculate the
minimum transverse field that can effectively withstand (and reverse) the
buoyant deformation (\cite{emfmi96}).  At this point, it is sufficient to
obtain an order of magnitude estimate for it that can serve as a guide for
the rest of the paper. This can be easily done for tubes with smooth
distributions like (\ref{eq_blong}) and (\ref{eq_pitchassympt}) or
(\ref{eq_pitchgauss}) and satisfying $\rtube^2 \ll \hp^2$. The resulting
criterion is best expressed in terms of the pitch angle calculated at the
position {\it of the maximum of the transverse field}, $\Psimax$; it reads:
\begin{equation}
\sin{\Psimax} \defdef \left(\frac{\Btran}{B}\right)_{\Max}^{} \gtrsim
\tilder^{1/2} \,
\left(\left|\frac{\drho}{\rho}\right|_0^{} \frac{\beta_0}{2}\right)^{1/2} \;.
\label{eq_min_pitch}
\end{equation}
For the tubes with initial condition \Sec\ref{sec_inicond}, this approximate
criterion should hold within about a factor $2$. A more precise condition
involving the detailed buoyancy distribution in the tube can be obtained by
studying the magnetostatic equilibrium of a horizontal twisted tube with a
non-homogeneous buoyancy distribution in the limit $\rtube^2 \ll \hp^2$
(\cite{emfmi96}). It can be shown (see their Eqs.~44 and 49) that for the
deformation of the tube to remain small, the radial profile of the pitch
angle, $\Psi(r)$, must satisfy:
\begin{equation}
\sin\Psi(r) \gtrsim \left(\frac{r}{\hp}\right)^{1/2} \left\{
\left|\frac{\drho(r)-\left<\drho\right>(r)}{\rho(r)}\right| \beta(r)
\right\}^{1/2}\;,
\label{eq_analytic_paper}  
\end{equation}
where $\left<f\right>(r)$ is the radial average of $f$ between the radii $0$
and $r$. At each radius $r$, the deformation of the magnetic field is
directly related to $\drho(r) - \left<\drho\right>(r)$, i.e., to the {\it
differential buoyancy} at that radius.

The threshold due to the differential buoyancy (Eq.~\ref{eq_min_pitch} or
\ref{eq_analytic_paper}), however, cannot be the only criterion of interest
for our problem.  In fact, even a uniformly buoyant tube, if untwisted, is
subjected to a deformation and conversion into vortex rolls.  A deforming
agent independent of the initial differential buoyancy is the pressure
profile (i.e., the pressure fluctuations, $\pextf$, above the background
stratification) built by the external flow around the boundary of the tube
(\cite{emfmi96}): a simple criterion for the resistance of the tube against
those fluctuations is
\begin{equation}
\frac{8\pi|\pextf|}{\Btran^2}\lesssim 1\;,
\label{eq_press_fluct}
\end{equation}
Now, $\pextf$ is itself of order the ram pressure of the external flow
relative to the tube. Thus, it is no higher than about $\rho\vterm{2}/2$.
Substituting from Eq.~(\ref{eq_term_speed_order}), we obtain again a 
criterion as in (\ref{eq_min_pitch}).  

%%%%%%%%%%%%%
\subsubsection{Vorticity generation}\label{sec_vorticity_eq}

Given the central role of the vorticity in the present problem, it is
adequate to understand the minimum condition (\ref{eq_min_pitch}) in terms
of the vorticity generation.  The time evolution of the longitudinal
component of the vorticity, $\vortlong$, is governed by the following
equation:
\begin{eqnarray}
\rho \frac{D}{Dt}\left(\frac{\vortlongvec}{\rho}\right) =
\nabla\left(\frac{\drho}{\rho}\right) \,\times\,\vec{g}
\;+\; \frac{\nabla\times\FLtranvec}{\rho}
\nonumber \\
\;+\;\nabla\left(\frac{1}{\rho}\right) \,\times\,\left[
-\nabla\left(\dpre + \frac{\Blong^2}{8\pi}\right) + \FLtranvec\right]\;.
\label{eq_vort}
\end{eqnarray}
\notetoeditor{The $\vortlongvec$ in equation(\ref{eq_vort}) is a VECTOR. It
must therefore be printed using bold faces like for the $\FLtranvec$.}
$\FLtranvec$ represents the projection of the Lorentz force on the transverse
plane $(x,z)$. For definiteness, when speaking about the sign of the
vorticity in the rest of this section, we refer to the right half of the tube
only.

At issue in this section is the generation of vorticity in the main body of
the tube (the vorticity in the tube periphery is discussed in
\ref{sec_vorticity_generation}). The first term on the right hand-side is the
counterpart in terms of vorticity of the gravitational torque: it produces
{\it positive} vorticity because $\nabla(\drho/\rho)$ points outwards.  The
second term on the right-hand side of (\ref{eq_vort}) represents the effect
of $\Btranvec$ and can be rewritten as:
\begin{equation}
\label{eq_vort_mag}
\frac{\nabla\times\FLtranvec}{\rho} =
\frac{\left(\Btranvec\cdot\nabla\right)\Jlongvec}{c\,\rho}\;,
\end{equation}
with $\Jlongvec$ the longitudinal component of the electric current density.
At the beginning of the run, the longitudinal current is axisymmetric so that
$\nabla\times\FLtranvec$ is zero. As soon as the tube center begins to rise
relative to the periphery, the longitudinal current is enhanced in the
upper half and diminishes in the lower half of the tube. According to
Eq.~(\ref{eq_vort_mag}), this produces {\it negative} vorticity in
the interior of the tube, i.e., it tends to counteract the effect of the
gravitational torque. By setting these two terms equal, we obtain a
criterion for the minimum transverse magnetic field that can effectively
oppose the initial deformation of the tube: the result is, again,
(\ref{eq_min_pitch}).  For the problem we are considering here, the last term
of the vorticity equation (\ref{eq_vort}) is O($\drho/\rho + L/\gamma\hp$)
smaller than the other two terms and is therefore not of primary interest
here ($L$ represents a local characteristic length for the transverse magnetic
field , e.g., the radius of the tube or the thickness of the tube boundary). 

Criteria equivalent to  (\ref{eq_min_pitch}) can also be obtained through other
physical considerations. For instance, Tsinganos (1980) obtained a similar
threshold for $\Psi$ by calculating the stability of the tube against
splitting due to the development of Rayleigh-Taylor and Kelvin-Helmholtz
instabilities at its apex.

%%%%%%%%%%%%%%
\subsection{Other parameters}\label{sec_other_params}

\subsubsection{Plasma beta and tube radius}\label{sec_beta_radius}

The choice of values for the parameters $\beta_0$ and $\tilder$ must be
guided both by (astro)physical insight and feasibility of the numerical
calculation.  $\beta_0$ is expected to be very high in the solar interior,
like, e.g., O($10^5$) for the magnetic tubes at the bottom of the convection
zone. The Courant condition for the numerical code, though, sets a stringent
upper limit to the value of $\beta_0$ that we can use. Using expression
(\ref{eq_term_speed_order}) we can calculate the number of timesteps
necessary for our tube to cross the whole integration box, $n_t$, as
\begin{equation}\label{eq_noof_timesteps}
n_t \gtrsim %\frac{\gamma}{\sqrt{\pi}\, 0.65} \,
n_z\left(\frac{\beta_0}{\tilder}\right)^{1/2} \;,
\end{equation}
with $n_z$ the total number of cells in the vertical direction. A trade-off
between field intensity and magnetic flux of the tube is then necessary.  As
a compromise we have chosen $\beta_0 = 10^3$ and $\tilder = 3.7\;10^{-2}$.
This should be sufficient to understand qualitatively the flows in and around
the tubes with the field strength and flux expected in the progenitors of
active regions.

\subsubsection{The Magnetic Reynolds number, $\Rm{}$}
\label{sec_magnreyn} 

The magnetic Reynolds number can be defined in this problem in terms of the
terminal speed of rise and the tube radius. Its order of magnitude is thus
given by $\Rm{} = \hbox{O}(\tilder^{3/2} / \tilde{\eta})$. The choice of a
value for this parameter must be guided primarily by numerical criteria:
large magnetic field gradients develop in the tube along the evolution,
specially at its upper rim. We have chosen the ohmic diffusivity so that the
resulting value of $\Rem{}$ is a few times $10^2$. This yields good numerical
performance but causes some unwanted diffusion of the field outward from the
tube. A measure for the latter can be obtained from the $y$-component of
Walen's equation, obtained by combining the continuity and induction
equations, (\ref{eq_continuity}) and (\ref{eq_induction}):
\begin{equation}\label{eq_walen}
\fracdps{D}{Dt}\left(\frac{\Blong}{\rho}\right) =
\left(\frac{\Bvec}{\rho}\cdot\nabla\right) v^y + 
\frac{\eta}{\rho}\,\Delta \Blong\;.
\end{equation}
In the central region of the tube, the variation in $\Blong/\rho$ is
basically due to diffusion. The calculations presented in the following
are not diffusion--dominated (as expected from the high $\Rem{}$), in the
sense that the decrease of this quantity is small in the tube center
during the period of time shown in the figures. This can also be checked
through an order of magnitude estimate: the diffusion term in
(\ref{eq_walen}) for the initial gaussian distribution can be compared to the
rate of change of the field (left-hand side of the same equation) imposed
by the rise of the tube. The ratio between them is
$(\Rem{}\,\tilder)^{-1}$, which is O($10^{-1}$) in the present case.

\subsubsection{The Weber number, $We$}
\label{sec_magnweber} 
In the hydrodynamical literature on air bubbles, 
\clearpage\onecolumn
\thispagestyle{empty}
\begin{figure}
\vskip \textheight
%\centerline{\psfig{figure=figure1.eps,height=1.0\textheight}}
%\plotfiddle{f1.eps}{0.5\vsize}{0}{30}{30}{-148}{0}
\caption{Rise of a twisted tube through a stratified environment. The
initial pitch angle $\Psimax$ is $7^\circ$. The grey scale and the contours
correspond to the longitudinal magnetic field intensity. At each time step
white corresponds to the maximum of $\Blong$ at this instant and black to 1
percent of this maximum. The arrows correspond to the velocity field. In the
figure, only a about 1/5 of the total integration box is reproduced; more
precisely, $100\,$x$\,500$ out of a total of $300\,$x$\,700$ points for each
half of the figure are included. The times represented are: $t=0$, $1.3$,
$2.5$, $3.8$, $5.3$, $6.6$. The complicated shape of the wake in the two last
panels is due to the episodic release of small secondary rolls from the
lateral edges of the tube (see \Sec6.2.3).\label{fig_rise}}
\end{figure}
\clearpage\twocolumn\noindent
the Weber number is used to measure the relative 
importance of the inertial forces of the flow to the
surface tension at the boundary of the bubble
(e.g. \cite{ryskinleal84b}). The role of surface tension is played in our
case by the jump in magnetic tension of the transverse component at the
boundary of the tube. Hence we define the magnetic Weber number as
\begin{equation}
We \defdef \frac{v^2 \rho}{\btran^2/(4\pi)}\;.
\label{eq_magn_weber}
\end{equation}
Inserting for $v$ the terminal speed we easily obtain $We =
\tilder\,\beta_0\,|\drho/\rho|_0^{}/(2 \sin^2\Psimax)$.  The condition on the
minimum pitch angle to avoid the splitting of the tube, (\ref{eq_min_pitch}),
can thus be reformulated as the condition that the {\it Weber number be at
most of order one}, \hbox{$We \lesssim 1$}.

%%%%%%%%%%%%%%%%%%%%%%%%%%%%%%%%%%%%%%%%%%%%%%%%%%%%%%%%%%%%%%%%%%%%%%

%%%%%%%%%%%%%%%%%%%%%%%%%%%%%%%%%%%%%%%%%%%%%%%%%%%%%%%%%%%%%%%%%%%%%%

\section{The rise of a twisted tube in a moderately stratified environment: a
representative case, $\Psimax=7^\circ$}
\label{sec_rise}
The critical value of the pitch angle parameter $\Psimax$ obtained by
substituting in (\ref{eq_min_pitch}) the chosen values of $\beta_0$,
$\tilder$ and $\Delta\rho/\rho$ is $6^\deg$.  In the present section we
describe some major features of the rise of a tube with a pitch angle
$\Psimax$ close to that value, $\Psimax=7^\circ$. The time evolution is
illustrated in Fig.~\ref{fig_rise}, which shows a fraction of the integration
box used ($30\%$ and $70\%$ of the box in the horizontal and vertical
directions, respectively). In \Sec\ref{sec_split} we compare the results for
flux tubes with different values of the initial $\Psimax$.

%%%%%%%%%%%
\subsection{The initial acceleration phase}
\label{sec_rise_accel}
In the absence of external flows to keep it in place, the tube lacks
equilibrium globally and starts rising, at the same time sending out sound
waves all across the box. The initial global acceleration is basically {\it
free-fall}, i.e., $(g/i)\; (\overline{\Delta\rho}/\overline{\rho})$, with an
overbar indicating average values in the tube and $i$ standing for the
enhanced inertia factor, which turns out to be $2$ (see
\Sec\ref{sec_term_velo}).

Superimposed on the global rise, different kinds of motions within and around
the tube take place which tend to deform its initial axisymmetric shape
(Fig.~\ref{fig_rise}, upper row).  One of them is the faster rise of the tube
center compared with the periphery due to the differential buoyancy
(\Sec\ref{sec_diff_buoy_pres_fluct}).  As a consequence of this, the magnetic
tension associated with the transverse field builds up at the tube front; the
relative motion of the tube center is thus stopped and reversed.  A vertical
oscillation of the tube center ensues within the tube's cross section.  The
compression/expansion in the upper/lower half of the tube together with the
internal oscillations are clearly visible in Fig.~\ref{fig_lm_center}, where
we have plotted the position as a function of time of several Lagrange
markers located along the vertical symmetry axis of the tube.

Simultaneously with the processes just explained, the external matter slides
around the tube and {\it drags} toward the rear the outermost tube layers
where the magnetic field is too weak to oppose any important Lorentz
force. The field is thus stretched all along the boundary
(\Sec\ref{sec_interface}). Vorticity is being generated in the matter being
dragged: as a result, two magnetized vortex rolls are created that trail the
tube motion (Fig.~\ref{fig_rise}, third panel). The main body of the tube
thereby loses about $30\%$ of its original magnetic flux. This figure
sensitively depends on the initial pitch angle (\Sec\ref{sec_split}). Given
the sign of the vorticity, the matter between the rolls is moving upwards
with respect to the back of the tube. A pressure excess appears at the upper
end of the inter-vortex space, directly below the lowermost tube layers.
\begin{figure}[t]
%\plotone{figure2.eps}
\vskip 6cm
\caption{Distance to the stagnation point at the tube front of $50$ mass
elements initially located at equidistant points along the vertical symmetry
axis as a function of time(solid lines). The compression of the tube front
and the oscillations of the central regions along the vertical axis can be
clearly seen. The curves plotted vertically (stars) are profiles of the
transverse field intensity, $\Btran$, along the vertical symmetry axis at
regular intervals of time.\label{fig_lm_center}}
\end{figure}

%%%%%%%%%%%
\subsection{Torsional oscillations of the tube interior}
\label{sec_oscillations}
\begin{figure}[t]
\plotone{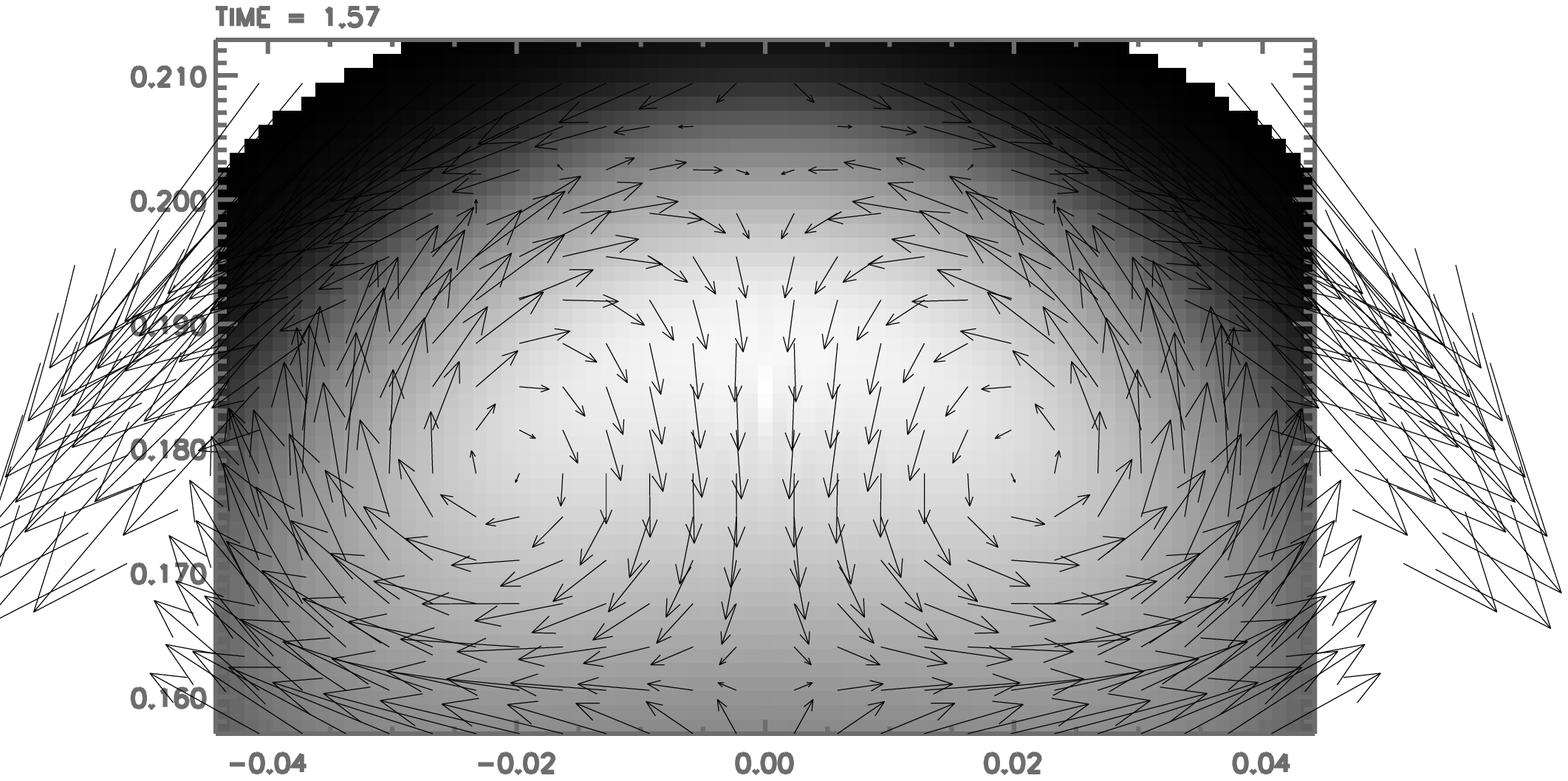}
\caption{Close-up view of the tube interior showing the velocity
  vectors of the internal torsional oscillation. The arrows in the figure
  correspond to the relative velocity of the individual mass elements with
  respect to the tube apex.\label{fig_torsional_oscill}}
\end{figure}

The vertical oscillations of the tube center are in fact part of a torsional
oscillation in which most of the tube interior is taking part
(Fig.~\ref{fig_torsional_oscill}). The torsional oscillation has left-right
symmetry: each tube half is rotating back and forth around a horizontal axis
offset by a fraction of $\rtube$ from the midplane. The energy of this
oscillation is being radiated away from the tube via pressure forces; it is
also damped through the diffusion (physical and numerical) present in the
code. Thus it slowly decreases in amplitude. The frequency of these
oscillations is of order $\omega_{\scs tors} = {\rm
O}[{v_A}_t/(2\pi\rtube)]$. Hence, one expects a few torsional oscillations to
be completed while the tube is rising across a scaleheight.

%%%%%%%%%%%%
\subsection{The asymptotic regime of rise}
\label{sec_asymp_reg}
The strong initial acceleration phase is followed by a quasi-stationary
asymptotic regime (Fig.~\ref{fig_rise}, lower row).  In it, the rate of
change of the tube interior and external flows becomes small compared with
the initial phases. The transition occurs at about $t=3$ (between the third
and fourth panels of Fig.~\ref{fig_rise}). The asymptotic regime is
characterized by the adoption of a terminal speed of rise, a sharp separation
of tube interior and surroundings and a well-developed trailing
wake. Simultaneously, the tube stops losing magnetic flux to the wake and
there is no further important deformation of the magnetic field lines in the
head of the tube. The following subsections deal with some of the features
characterizing this asymptotic regime.

\subsubsection{The terminal velocity}\label{sec_term_velo}
Once the wake is developed, the resistance of the surroundings to the advance
of the tube can be calculated using the customary aerodynamic drag force with
almost constant drag coefficient $C_D$. In the asymptotic regime, the total
buoyancy of the tube is a slowly varying quantity.  Thus we can expect the
tube to reach a terminal speed of rise given by:
\begin{equation}
\vterm{} =  \left(\frac{2\,g \,\csarea\,\overline{\drho} }{\CD
\csdiam\,\rho_e } \right)^{1/2}
 \,, \label{eq_term_speed_exact}
\end{equation}
$\csarea$ and $\csdiam$ being the cross sectional area and horizontal
diameter, respectively. The order of magnitude of this speed was estimated in
Eq.~(\ref{eq_term_speed_order}). The best fit to the numerical results yields
a value of $\CD$ of approximately $1.6$. The approach to this asymptotic
regime is not immediate to formulate analytically, since the coefficient
$C_D$ is not constant while the tube is changing its shape and developing a
wake. The initial acceleration of the tube as a whole is
\begin{equation}
a = \frac{g \,\overline{\drho} }{i\, \overline{\rho}} 
\,; 
\label{eq_accel_vterm}
\end{equation}
the factor $i$ allows for the added inertia due to the co-acceleration of the
external medium ($i=2$ for a rigid straight cylinder). To approximate the
velocity of the tube in the initial and intermediate time-dependent phases,
we use the expression for a tube which starts from rest with acceleration
$a$, is being acted upon by a constant driving force and has fixed $C_D$,
viz.:
\begin{figure}[H]
\plotone{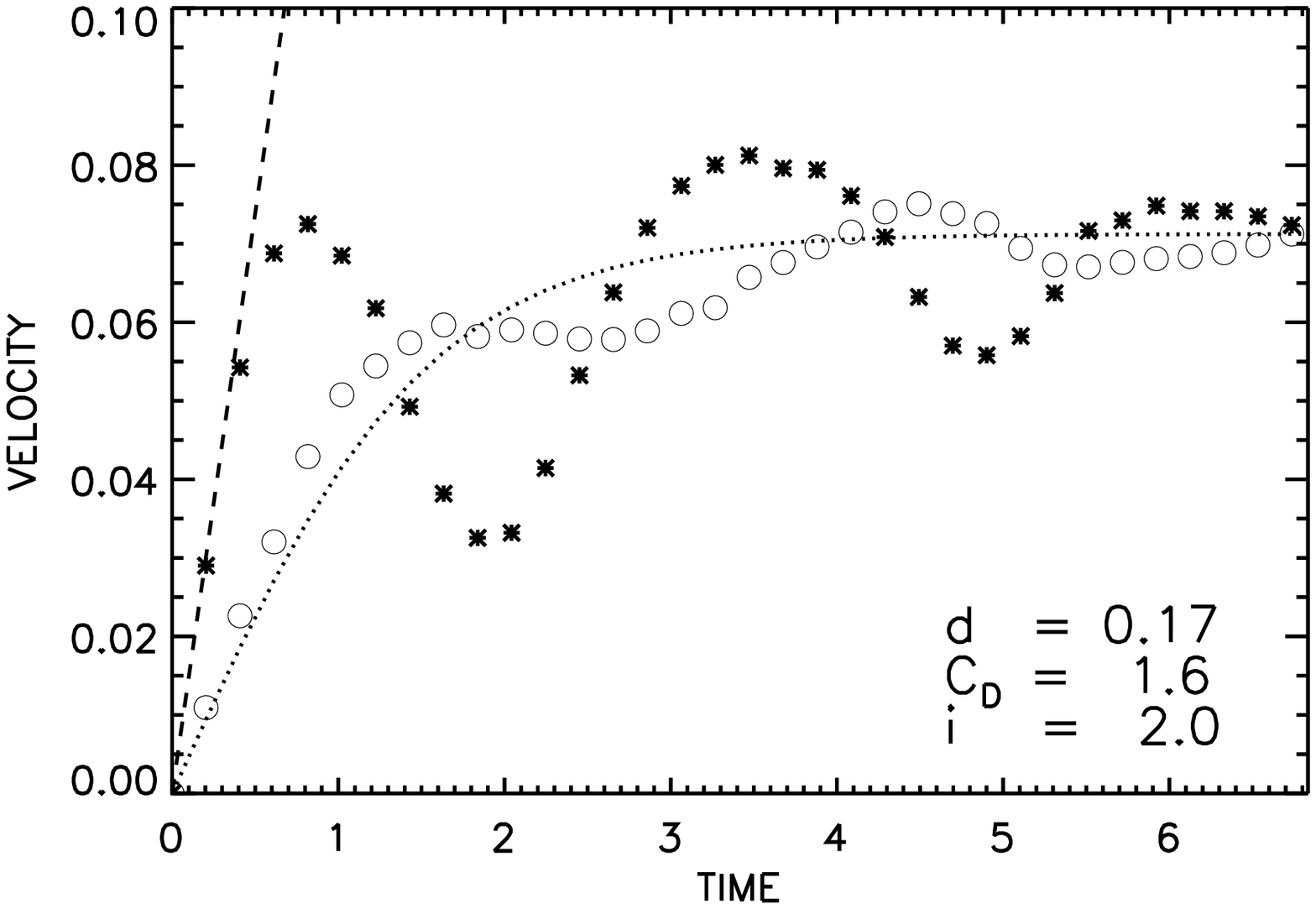}
\caption{Velocity of the tube center (stars) and of the apex
(circles) as a function of time. The dotted line corresponds to Eq.~(21). The
dashed line represents the initial acceleration of the tube center,
$(|\Delta\rho/\rho|)_{r=0}\;g/i$, with $i=2$. 
\label{fig_velocity}} 
\end{figure}
\noindent a constant driving force and has fixed $C_D$, viz.:
\begin{equation}
\vrise{} = \vterm{} \, \tanh \left( \frac{a\; t}{\vterm{}}\right) \;, 
\label{eq_velocity}
\end{equation}
with $\vterm{}$ still given by Eq.~(\ref{eq_term_speed_exact}).

The numerical results show a surprising closeness to Eq.~(\ref{eq_velocity}).
In Fig.~\ref{fig_velocity}, we show with a dotted line the velocity given by
Eq.~(\ref{eq_velocity}), while the actual speed of the tube apex, $\vapex{}$,
is indicated 
as circles. As can be seen, the tube speed {\it oscillates} around the mean
speed given by Eq.~(\ref{eq_velocity}). This is due to the strong
oscillations of the tube interior: in fact, the tube center oscillates with a
much larger amplitude, as is shown by the curve with asterisks. The tube
center {\it pushes} the apex, so that circles and asterisks are slightly out
of phase. The amplitude of the oscillation diminishes as its energy is being
radiated to the external medium and both curves converge toward the dotted
line. The initial acceleration of the tube apex is very close to
(\ref{eq_accel_vterm}) with $i=2$, i.e., as corresponds to the global
buo-
\begin{figure}[H]
\vskip 8.5cm
%\plotone{figure5.eps}
\caption{Equipartition line at $t=3.8$ together with the flow field relative
to the apex of the tube. The simple structure of the wake at this time is
clearly visible. The envelope of tube and wake has a well defined elliptical
shape. This figure bears a strong resemblance to the experimental results of
air bubbles rising in liquids (e.g., Collins 1965, also reproduced by
Batchelor 1967, plate 15).\label{fig_equipartition_line}}
\end{figure}
\noindent yancy of the tube. The tube center, in turn, has a higher
acceleration, corresponding to the local value of the buoyancy, namely
$(|\Delta\rho/\rho|)_{r=0}\;g/i$ (this value is shown in the figure as a
dashed line). Here again, the enhanced inertia factor $i$ must be set equal
$2$. 

As a test to the validity of the terminal velocity formula, we have
calculated the evolution of tubes starting with the same initial condition as
the one used to obtain Fig.~\ref{fig_velocity} (in particular, the same tube
radius, $\rtube$) but with $\beta_0$ chosen in a range between $1000$ and
$100$. The velocity curves of all these tubes are similar to those shown in
Fig.~\ref{fig_velocity}, with the time scale contracted by a factor
$\beta_0^{1/2}$ and the velocity scale enlarged by the same factor. More
precisely, if all these tubes reach the aerodynamic drag regime with terminal
velocity given by Eq.~(\ref{eq_term_speed_exact}), then the curves depicting
the velocity of the apex normalized by the plasma beta, $\vapex{}
\beta_0^{1/2}$, versus the position of the apex must all basically coincide
(the drag coefficient $C_D$ does not change much with varying Reynolds number
in this regime). The results of the tests show indeed a large degree of
superposition of those curves: the maximum relative deviation between them is
a factor $1.1$, which can be attributed to the change in $C_D$ consequent
with the variation in the Reynolds number.

\subsubsection{The equipartition line at the tube periphery}
\label{sec_equip_line} 
As a result of the resistance to deformation provided by the transverse
field, there is a neat separation between the tube interior and the outside
in the asymptotic regime. A good indicator of the location of the tube
boundary can be obtained by comparing the kinetic energy density of the
relative flow field, $\vreltvec\defdef\vec{v}-\vapexvec$, viz.~$\ekin \defdef
\rho \vrelt{2}/2$, with the energy density of the transverse magnetic field,
$\etran \defdef \Btran^2/(8\pi)$. We define the {\it equipartition line} as
the locus of those points in the tube periphery where the two energy
densities are equal.  Fig.~\ref{fig_equipartition_line} shows, at time
$t=3.8$, the relative flow field $\vreltvec$ and, superimposed on it, the
equipartition line. The flow field changes markedly when going across the
equipartition line: the velocities inside are much smaller than outside
it. In other words, once the terminal velocity has been reached the tube
rises basically as a unity, with only a weak internal flow pattern
corresponding mainly to the torsional oscillations described in the previous
section. A comparison of Fig.~\ref{fig_equipartition_line} with the results
of laboratory experiments (e.g., \cite{collins65} 
\begin{figure}[t]
\plotone{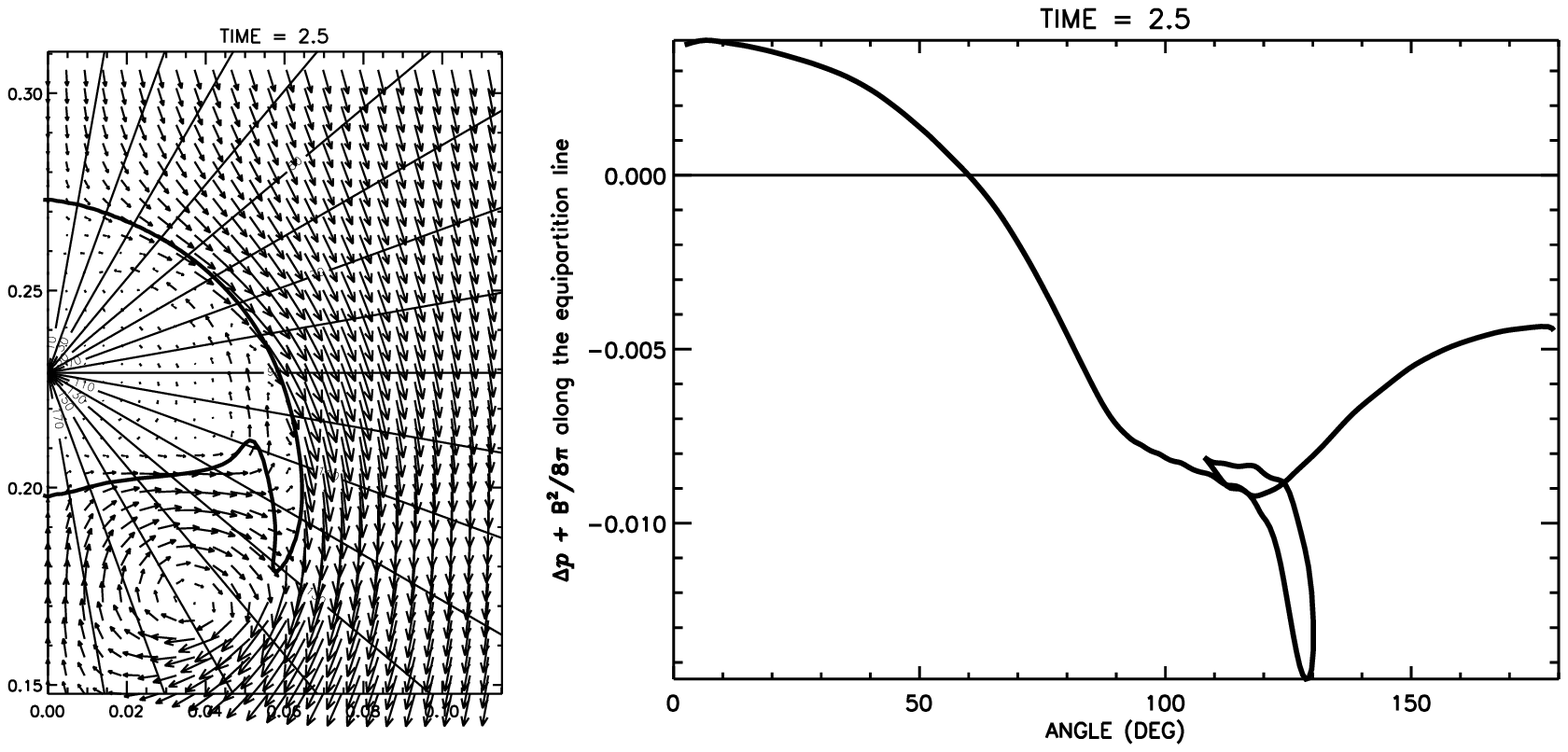}
\caption{Profile of the total pressure excess along the boundary of the tube,
which is defined by the equipartition line (see \Sec4.3.2), at time
$t=2.5$. The boundary of the tube (the equipartition line) is shown as a
thick line in the left panel. The feature in the middle of the profile
(between $110^\circ$ and $130^\circ$) results from the shape of the
equipartition line at the lateral edge of the tube. The arrows on the left
panel correspond to $\vreltvec$ and the radial lines are drawn at constant
azimuthal angles. \label{fig_bound_press}}
\end{figure}
or plate 15 in
\cite{batchelor67}), shows that there are striking similarities between
buoyant magnetic tubes and air bubbles rising in a liquid, both in the shape
of the rising object itself as in the wake, in spite of the very different
parameter values involved (like, e.g., the density deficit).

All along the initial phase, a pressure profile is set up around the tube
periphery which closely resembles the pressure distribution around rigid
cylinders in relative motion with the surroundings. To show this, we can plot
(Fig.~\ref{fig_bound_press}) the pressure in the points along the
equipartition line. In this figure, the pressure excess associated with the
return flow between the vortex rolls of the wake is clearly visible.
%%%%%%%%%%%%%%%%%%%%%%%%%%%%%%%%%%%%%%%%%%%%%%%%%%%%%%%%%%%%%%%%%%%%%%
%%%%%%%%%%%%%%%%%%%%%%%%%%%%%%%%%%%%%%%%%%%%%%%%%%%%%%%%%%%%%%%%%%%%%%

%%%%%%%%%%%%%%%%%%%%%%%%%%%%%%%%%%%%%%%%%%%%%%%%%%%%%%%%%%%%%%%%%%%%%%
%%%%%%%%%%%%%%%%%%%%%%%%%%%%%%%%%%%%%%%%%%%%%%%%%%%%%%%%%%%%%%%%%%%%%%
\section{The sharp interface at the tube boundary: magnetic boundary layer} 
\label{sec_interface}
All around the equipartition line there is a region with non-zero
vorticity. This contains (1) a boundary layer surrounding the tube and (2)
the trailing wake.Further out, there is the external medium with a largely
vorticity-free flow. In this section we study the structure of the boundary
layer; the wake is considered in \Sec\ref{sec_wake}.

The two main physical features characterizing the magnetic boundary layer
around the tube are a strong shear of the tangential flow and high magnetic
field gradients. They lead to enhanced generation of vorticity, ohmic
diffusion and generation of entropy.  In the following subsections we study
these items in turn:

%%%%%%%%%%%%
\subsection{Shear flow and field gradients}\label{sec_shearflow}
A pronounced shear is taking place in a thin band around the equipartition
line (Fig.~\ref{fig_equipartition_line}). As a result, the matter elements
and the transverse field are being stretched at the tube periphery all around
the tube.  To visualize this phenomenon, we use {\it Lagrange markers} which
follow the motion of the tube's mass elements (Fig.~\ref{fig_lagrange}).  We
choose six groups of markers at time $t=0$ (left panel), at a distance of the
tube center such that $\Blong$ is $5$\% (white asterisks), $27$\% (black
triangles) and $77$\% (black diamonds), respectively, of $\Blongmax$ (the
maximum value of $\Blong$ in the tube) both in the upper and lower half of
the tube. The markers are set in the neighborhood of the tube axis and are
left to evolve with the tube. As time advances, the markers of the outermost
group in the upper tube half are stretched by a very large factor along the
tube periphery (center and right panel of the figure); in fact, many of them
are brought all the way down to the wake. The markers of the group
immediately below them (triangles) are brought close to the uppermost group
and stretched, albeit by a smaller factor than the other group. Large
compression and stretching happen basically only in the neighborhood of the
interface between tube and surroundings: closer to the tube center (diamonds)
the mass elements are just periodically moved following the internal
torsional oscillation. At the rear of the tube, the Lagrange markers with
$\Blong = 0.27 \, \Blongmax $ (triangles) are also somewhat stretched by the
flow of the trailing wake (see \Sec\ref{sec_wake}). The outermost parcels at
the back of the tube (white asterisks) are trapped in the tail and no longer
rise with the tube.
%\newpage
\begin{figure}[t]
\vskip 0.45\textheight
%{\centering \leavevmode \psfig{figure=figure7.eps,width=1.0\textwidth}}
\hbox{\vtop{\hsize=\textwidth \caption{Motion of individual mass elements in
the tube, showing the stretching of the matter and field along the tube
boundary. The panels correspond to $t=0$, $1.9$ and $3.8$. At time zero the
Lagrange markers are located at $R/\hp=0.044$ and $0.068$ where $\Blong$ is
equal to $27$\% and $5$\% of its value at the center. Radial compression,
azimuthal stretching and shearing of the upper layers of the tube is
apparent. The elements indicated by black triangles at the back are also
somewhat stretched as a result of flow in the wake.\label{fig_lagrange}}}}
\end{figure}

\begin{figure}[H]
\plotone{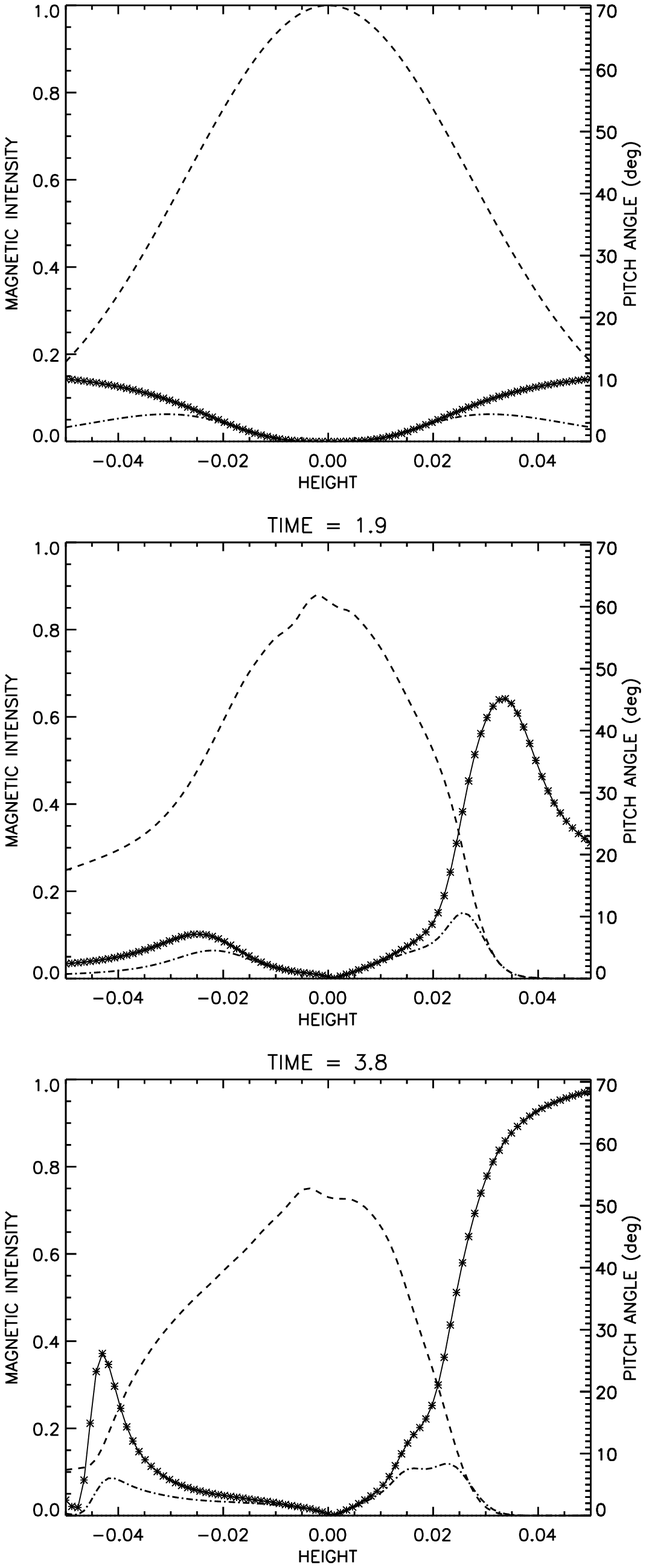}
\caption{Pitch angle distribution (stars) along the vertical
  central axis at three different times corresponding to the 3 panels of
  Fig.~7. The profiles of $\Blong$ (dashed) and $\Btran$ (dotted-dashed) are
  overplotted.\label{fig_pitch_angle}}
\end{figure}

Corresponding to this shearing, the pitch angle increases to large values at
the tube periphery: in Fig.~\ref{fig_pitch_angle} we plot the pitch angle
distribution along the vertical axis of symmetry for the three instants shown
in Fig.~\ref{fig_lagrange}.  The pitch angle becomes large in the interface
at the tube front (middle panel). In the quasi-stationary regime, its
distribution shows a horizontal asymptote outside the tube similar to the one
of the initial condition. This feature is {\it not} a mere consequence of our
initial condition (\ref{eq_pitchassympt}). In fact, an initial pitch angle
distribution with exponential decline for large radius at time $t=0$ (see
Eq.~\ref{eq_pitchgauss}) yields in the quasi-stationary regime a horizontal
asymptote similar to the one of Fig.~\ref{fig_pitch_angle}. It is then
perhaps more natural to choose the initial condition of
Eq.~(\ref{eq_pitchassympt}) from the outset.

The large pitch angles apparent on the right of Fig.~\ref{fig_pitch_angle}
are all in a region of the tube where the magnetic field intensity is very
low; the central regions, on the other hand, have low pitch angles. Thus, one
should not expect a global kink-unstable behavior of the tube. This can be
checked by calculating the net tension in the tube over background (see
\Sec9.2 in \cite{parker79}). With the convention that tension is positive and
pressure negative, we obtain a positive value indicating that the tube is
under longitudinal tension rather than compression. Thus, there should be no
tendency to buckling of the tube.
%
%%%%%%%%%%%%%%%%%
\subsection{Vorticity generation in the boundary layer}
\label{sec_vorticity_generation} 

The large jump in tangential speed visible across the equipartition line in
Fig.~\ref{fig_equipartition_line} marks the presence of a {\it vortex
sheet} surrounding the tube. In fact, vort-
\begin{figure}[t]
\vskip 12.2truecm
\end{figure}
\noindent icity (more precisely, its longitudinal 
component $\vortl$) is generated at the boundary layer during the whole
duration of the run. From there, it is advected toward the wake. 

These processes are governed by Eq.~(\ref{eq_vort}). As in
\Sec\ref{sec_tube_unity}, the right-hand side is dominated by the first two
terms [the third is again O($1/\beta_0$) smaller]. However, along the front
of the tube those two terms, i.e., the gravitational and magnetic vorticity
sources, {\it reinforce each other} rather than mutually cancel, as in the
tube interior. This is because the radial derivative of the transverse field
component reverses its sign in the neighborhood of the equipartition
line. For instance, in the simple case in which the field lines are still not
far from circular (e.g., along the front) the magnetic contribution is
approximately, 
\begin{equation}
\label{eq_current}
\frac{\nabla\times\FLtranvec}{\rho} \cong 
\frac{\Btran}{c\,\rho}\,\frac{\partial}{\partial\phi} \underbrace{\left[
\frac{1}{r} \frac{\partial}{\partial r} \left(r\Btran\right)
\right]}_{\cong\Jlong}\;,
\end{equation}
\noindent so that $\Jlong$ changes sign at about the maximum of $\Btran$.  As
a result, 
(and taking into account the sign of the azimuthal derivative
$\partial/\partial\phi$), in the right half of the tube positive vorticity is
generated all along the front of the tube, and negative vorticity is
generated at the rear, i.e., at the interface between the tube and the
wake. At the rear $\nabla(\drho/\rho)$ is approximately parallel to $\vec{g}$
and therefore the magnetic stress is anyway the main source of vorticity
there.

\begin{figure}[H]
\plotone{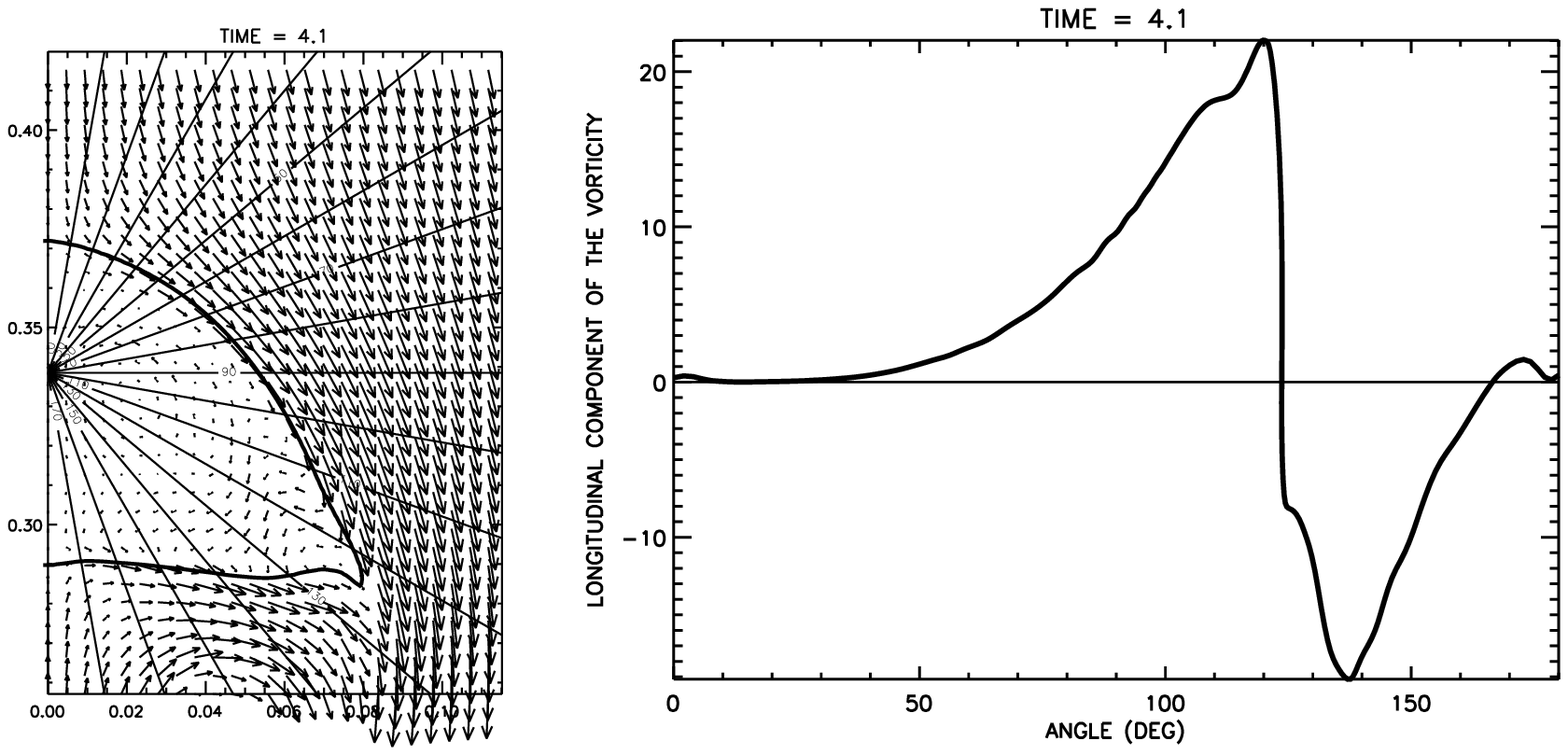}
\caption{Profile of the longitudinal component of the vorticity
  (right panel) along the equipartition line drawn in the left panel for
  $t=3.8$. The zero crossing at about $120^\circ$ corresponds to the
  transition from tube front to rear at the edge of the tube. The radial
  lines in the left panel are drawn at constant azimuthal
  angles.\label{fig_vort_angle_12}}
\end{figure}

In Fig.~\ref{fig_vorticity_halo} we show the distribution of vorticity in the
box at four different times during the evolution.  White and black indicate
positive and negative vorticity, respectively (i.e., clockwise and
anticlockwise rotation), whereas the grey background is the zero vorticity
level. Soon after the beginning (leftmost panel; $t=0.3$) the gravitational
torque is creating positive and negative vorticity in the right and left tube
halves, respectively, across the whole tube interior. In the second panel
($t=1.6$), this has already been countered by the magnetic tension of the
transverse field: the reversal of the sign of the vorticity in the interior
reveals the first backward torsional oscillation. Additionally, a vortex
sheet is already apparent all around the tube front. At the rear, there is
only an incipient wake and no clear vortex sheet yet. The third and fourth
panels correspond to the asymptotic phase ($t=3.8$, i.e., close to the
beginning of the stationary regime, and at an advanced stage, $t=5$,
respectively).  Vorticity generation now occurs basically close to the
equipartition line only. Vortex sheets along {\it both} the front and the
rear of the head of the tube are visible. As expected, the sign of the
vorticity in them is 
opposite. A detailed distribution of $\omega_l$ in the boundary layer around
the tube (from top to bottom), can be seen in Fig.~\ref{fig_vort_angle_12}.
The vorticity indeed changes sign at the edge of the tube: the abrupt shape
of the zero crossing is due to the acute shape of the latter. The small
maximum at the right end of the distribution is characteristic of the tail
(\Sec\ref{sec_wake_structure}).
\begin{figure}[t]
\vskip 0.41\textheight
%{\centering \leavevmode \psfig{figure=figure9.eps,width=1.0\textwidth}}
\hbox{\vtop{\hsize=\textwidth \caption{Distribution of the $y$--component of
the vorticity 
  vector in and around the tube at $t=0.3$, $1.6$, $3.8$ and $5.0$. White
  corresponds to positive vorticity (clockwise rotation) and black to the
  opposite. The two leftmost panels clearly show the vorticity associated
  with the torsional oscillations. The boundary layer and vortex sheet at the
  tube front are visible in the three rightmost panels. In the two last
  panels the vortex sheet at the interface between the wake and the tube is
  also visible.
\label{fig_vorticity_halo}}   }}
\end{figure}

%%%%%%%%%%%%%%%%
\subsection{Ohmic diffusion and field advection}\label{sec_ohmdiff_fadv} 

\begin{figure}[t]
\plotone{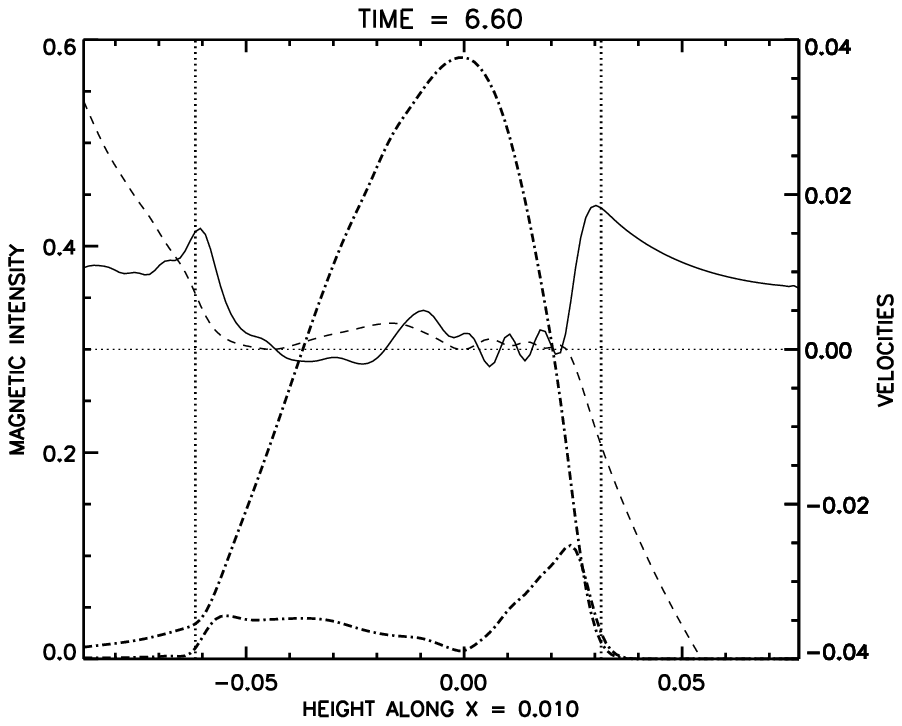}
\caption{Plots of $\Blong$ and $\Btran$ (dash-dotted) together
  with $v_x$ (solid line) and $(\vrel{})_z$ (dashed) along a vertical axis at
  $x=0.01$. The vertical dotted lines correspond to the positions where
  $\etran=\ekin$.\label{fig_velo_interface}}
\end{figure}

The structure of the magnetic boundary layer is dominated by the ohmic
diffusivity, which in part plays in the present case the role of the
viscosity in the standard hydrodynamic boundary layers. An important
difference between viscous and purely magnetic boundary layers, though,
concerns the no-slip condition for the flow: in the purely magnetic case,
there is no counterpart to the viscous second-order derivative term that
could force the tangential velocity to reach a zero value at the boundary. In
our case, however, the flows on the inner side of the equipartition line are
reduced or suppressed by the magnetic forces.  In other words, in the present
boundary layer the Lorentz force imposes a {\it no-slip} condition which
does not allow a penetration of the external incoming flow into the tube nor
large fluctuations of the transverse velocity inside.  This is clearly
visible in Fig.~\ref{fig_velo_interface} where we have plotted $v_x$,
$(\vrel{})_z$, $\Btran$ and $\Blong$ along a vertical axis slightly offset
from the symmetry axis.

The slope of the profile of $\Btran$ and $\Blong$ across the boundary
layer results from the equilibrium between the outward 
ohmic diffusion and the field advection through the external flow. For
$\Btran$, for instance, this is
controlled by the transverse component of the induction equation
(Eq.~\ref{eq_induction}). Written in terms of the magnetic potential $A$,
the latter reads: 
\begin{equation}
\left(\vreltvec\cdot\nabla\right) A = \eta \;\Delta A\;,
\quad \hbox{with}
\quad \Btranvec  \defdef \nabla \times(A\vec{i}_y)\;.
\label{eq_induction_y}
\end{equation}
As the transverse field diffuses outward, it {\it swallows} 
\begin{figure}[t]
\vskip 11.22truecm
\end{figure}
\noindent some of the
incoming external matter (which, in this way, slowly enters the magnetic
field system). Simultaneously, this same external matter {\it advects} the
magnetic field back toward the tube interior. The equilibrium between
those two terms is reached for a given thickness of the boundary layer,
$\Lbl$. To obtain an estimate for $\Lbl$, we note that the external flow
has a lengthscale comparable with the tube radius (a common feature of
cylinders moving in fluid media), so that,
\begin{equation}
\Lbl \sim \fracdps{R}{\Rem{1/2}}\;.
\label{eq_bl_thickness}
\end{equation}
This estimate is in agreement within a factor about $2$ with the results
of the numerical calculation (see Fig.~\ref{fig_velo_interface}).
An analogous advection--diffusion equilibrium can be seen to hold for
$\Blong$ in the boundary layer.
%

%%%%%%%%%%%%%%%
\subsection{Comparison with air bubbles and rigid tubes.}
\label{sec_bound_bubbles}

In spite of the striking similarities between the present problem and the
rise of an air bubble, the corresponding boundary layers at the periphery of
the rising object differ in important ways. The boundary of an air bubble is
basically a {\it free surface} for the external medium, since the density and
viscosity in the interior are negligible compared with those outside. The
external flow must then satisfy a zero tangential stress boundary condition
rather than the no-slip condition of the present magnetic tube (e.g.,
\cite{batchelor67}, \Sec5.14; \cite{ryskinleal84a}, 1984b). As a consequence,
the tangential velocity at the gas/liquid interface in the air bubble does
not vanish, but must instead satisfy the following equation:
\begin{equation}
\omega_l = 2 \,\kappa\, (\vrel{})_{tg}^{} \;,
\label{eq_vort_bubble}
\end{equation}
with $\kappa$ the curvature of the boundary layer and $(\vrel{})_{tg}^{}$ the
tangential component of the relative flow speed.  The resulting jump in
tangential velocity across the boundary layer of a bubble is therefore only
of order $Re^{-1/2}$, while it is O($1$) for a no-slip boundary layer, as in
the present problem.

The magnetized boundary layer of the rising tube, in fact, resembles more
closely the boundary layers around solid bodies with a no-slip condition than
those around air bubbles. The distribution of vorticity along the
equipartition line shown in Fig.~\ref{fig_vort_angle_12} is quite similar to
that around the boundary of a rigid cylinder at $Re\cong 300$ (e.g.,
\cite{taphuoc80}). In our case, though, the reversal of the sign occurs more
abruptly because of the particular geometry of the equipartition line.

%%%%%%%%%%%%%%
\subsection{Entropy generation}\label{sec_entr_gener}
The enhanced ohmic dissipation in the boundary layer is creating entropy
at a rate:
\begin{equation}
\frac{D}{Dt}\left(\frac{\Delta s}{\cp}\right) = \eta \frac{\gamma-1}{\gamma}
\, \frac{\left(\nabla\times\vec{B}\right)^2}{4\pi p} \;,
\label{eq_entropy}
\end{equation}
To order of magnitude, this entropy increase can be written:
\begin{equation}
\frac{D}{Dt}\left(\frac{\Delta s}{\cp}\right) \approx
\frac{1}{\betabl}  \frac{\vrise{}}{R} 
\;, \label{eq_entropy_magn}
\end{equation}
with $\betabl$ the local plasma beta in the boundary layer.  The entropy
increase for a matter element moving all along the tube boundary is then of
order $1/\betabl$. As a result, the density difference $\Delta\rho/\rho$ of
those elements can be substantially modified.  This, however, is unlikely to
cause important modifications in the dynamical behavior of the tube
boundary. The main driving force there, the pressure gradient, is of order
$\rho\vrise{2}/R$; therefore, it should be more important than the local
gravitational term $g\Delta\rho$ by the ratio $B_0^2/\Bbl{2} \gg 1$, where
$\Bbl{}$ is the magnetic field strength in the boundary layer.

In our numerical simulation we also have some increase of entropy due to the
artificial viscosity. Its influence, however is only secondary because of the
low value of $\Rm{}$ chosen. 

%%%%%%%%%%%%%%%%%%%%%%%%%%%%%%%%%%%%%%%%%%%%%%%%%%%%%%%%%%%%%%%%%%%%%%
%%%%%%%%%%%%%%%%%%%%%%%%%%%%%%%%%%%%%%%%%%%%%%%%%%%%%%%%%%%%%%%%%%%%%%
\section{The wake}
\label{sec_wake}
Wakes have been the object of active research in fluid dynamics for at least
the past four decades. This includes the wake behind solid cylinders (e.g.,
\cite{collinsdennis73a}; \cite{collinsdennis73b}; \cite{taphuoc80},
\cite{bouardcoutanceau80}; \cite{taphuocbouard85}), drops
(\cite{dandyleal89}; \cite{stone94}) and air bubbles
(\cite{daviestaylor50}; 
\cite{collins65}; \cite{parlange69}; \cite{wegenerparlange73};
\cite{hnatbuck76}; \cite{ryskinleal84b}; \cite{christovvolkov85}). In
this chapter we briefly describe the formation and 
structure of the wake behind the rising magnetic tube and pay special
attention to the similarities and differences to other wakes known in the
literature.

%%%%%%%%%%%%%
\subsection{The time evolution of the wake}
\label{sec_wake_evolution}

Solid cylinders, air bubbles and drops have a clearly defined boundary from
the beginning. Vorticity is created along the boundary and advected
downflow. The wake is formed out of external fluid only, via, e.g. in solid
bodies, the detachment of the boundary layer. In our case, in contrast, there
is no such clear boundary at the beginning. The layers of weak magnetic field
in the outskirts of the tube (a) are unable to resist the incipient external
flow so that they are bodily convected to the rear and (b) increase their
vorticity through the mechanisms explained in \Sec\ref{sec_vorticity_eq} and
\Sec\ref{sec_vorticity_generation}. It is the accumulation of this rotating
material at the back of the tube that constitutes the wake. The material
making up the wake may in fact be a large fraction of the initial tube in the
case of small initial pitch angles (\Sec\ref{sec_split}).

\begin{figure}[t]
\plotone{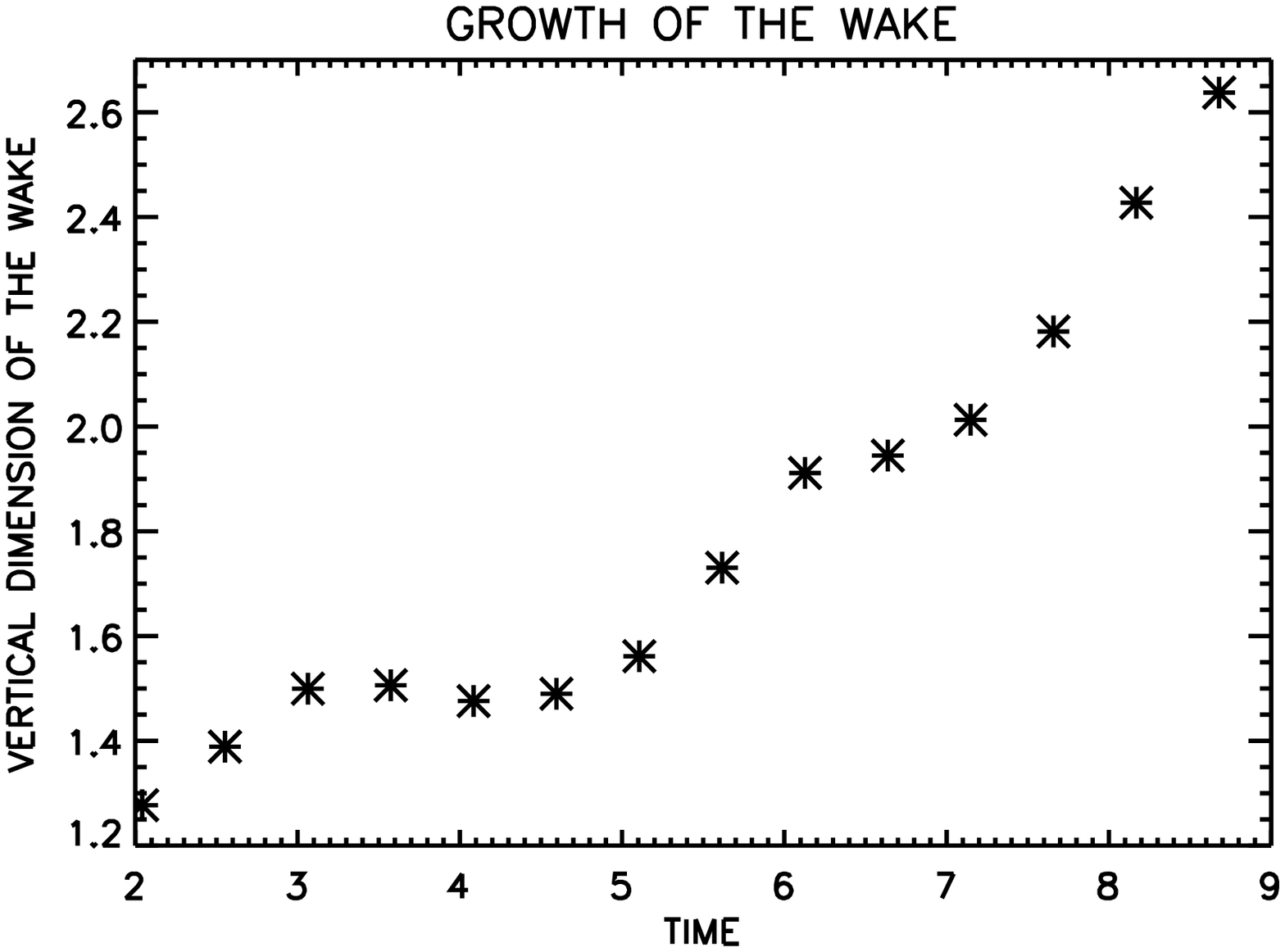}
\caption{Vertical size of the wake as a function of time,
  calculated as the distance between the stagnation point at the rear of the
  tube and the stagnation point at the bottom of the wake. The growth is
  quasi-linear, with oscillations superimposed.\label{fig_wake_growth}}
\end{figure}

The process of formation of the wake is basically complete by $t=3.0$ (see
Figs.~\ref{fig_rise} and \ref{fig_equipartition_line}) and this signals the
transition to the asymptotic regime of rise (\Sec\ref{sec_asymp_reg}). The
latter, however, is not exactly stationary, since the wake does not remain
unmodified in the rest of the evolution. In fact, it slowly grows in size and
episodically changes its shape. The growth can be seen in
Fig.~\ref{fig_wake_growth}, where the vertical size of the rolls is depicted
as a function of time.  The elongation is roughly linear, with a
small-amplitude oscillation superimposed. This kind of behavior is not
unknown in fluid dynamics. The elongation of the wake is indeed similar to
what is observed behind a rigid cylinder moving at constant speed with
$R_e\cong 300$ after an impulsive start (see \cite{taphuoc80}): in
the laboratory experiment the linear growth of the wake with time continues
well after the high-$R_e$ flows have been set up around the cylinder and $C_D$
has reached a constant value close to $1$. For higher values of $R_e$
($=3000$) the growth of the wake with time is exponential rather than linear
(\cite{taphuocbouard85}).  In our case, a small part of the growth of the
wake is due to the stratification; however, its main cause is the transport
of rotating material from the boundary layer along the boundary of the tube
to the wake.  This is clearly visible in the third and fourth panels of
Fig.~\ref{fig_vorticity_halo}: a white {\it streamer} of positive vorticity
coming from the tube front is being wrapped up first around and then into the
roll on the right-hand side of the wake. Simultaneously, a black {\it tongue}
with origin in the back of the tube is also making inroads into the wake,
next and following a parallel path to the white streamer.

In the case of rigid tubes and bubbles moving through a fluid (and assuming
perfect symmetry about the object's midplane), the growth of the wake
normally lasts until a dynamical equilibrium is adopted.  In it, the
vorticity generated in the boundary layer is advected to the wake, where it
is redistributed through viscous diffusion and further
advection. Simultaneously, the energy gained through the driving force (the
potential energy for the bubbles) is transformed into heat in the wake, again
through viscous dissipation (\cite{parlange69}; \cite{wegenerparlange73}). 
Further changes may occur if there is no mechanism to dispose of the heat
thus generated. 
\begin{figure}[t]
\vskip 0.35\textheight
%{\centering \leavevmode \psfig{figure=figure13.eps,width=1.0\textwidth}}
\hbox{\vtop{\hsize=\textwidth \caption{Formation and release of a secondary roll at the edge of
  the tube. The panels contain a time series showing the equipartition line
  and the relative flow field. The secondary roll is being released to the
  downflow; later on, it gets dragged all around the wake and noticeably
  perturbs the shape and velocity distribution of the latter. Note also the
  jumps in velocity inside the main roll of the wake, revealing the presence
  of vortex sheets which have been advected from the tube front.
  \label{fig_second_roll}} }}
\end{figure}

In the results presented here, no dynamical equilibrium is reached since
there is no equivalent mechanism to diffuse the vorticity effectively. This
can be seen quantitatively as follows: the characteristic time for the
adoption of a stationary state in the wake through viscous diffusion and
dissipation is $Re\; (\Lwake/\rtube)^2$ times longer than the characteristic
time for the formation of the wake itself, $\Lwake$ being the transverse
dimension of the wake and $Re$ the viscous Reynolds number. In our case,
$\Lwake \approx \rtube$ and we expect $Re$ (e.g., calculated on the basis of
the numerical viscosity) to be at least as high as $\Rm{}$. Therefore, the
establishment of the dynamical equilibrium lasts much longer than the
evolution presented in this paper. The consequence of this is the very
inhomogeneous vorticity distribution apparent in
Fig.~\ref{fig_vorticity_halo}. In a high-$Re$ case, in fact, it may never
come to a full dynamical equilibrium because an instability interrupts the
approach to a steady state: the two standing rolls develop asymmetric
oscillations until one of them is released downstream (see, e.g.,
\cite{batchelor67}, plate $10$). In contrast, in low-$Re$ bubble experiments,
the trailing vortex rolls are similar to a Hill's vortex (e.g.,
\cite{collins65}; \cite{parlange69}; \cite{wegenerparlange73};
\cite{ryskinleal84b}; \cite{christovvolkov85}).
\begin{figure}[t]
\vskip 9.8truecm
\end{figure}
%
   
%%%%%%%%%%%%
\subsection{Further structural features common with laboratory experiments.}
\label{sec_wake_structure}

%%%%%%%%%%%%%%%
\subsubsection{The skirt}
\label{sec_skirt}

In experiments with buoyant air bubbles, it is often observed that a small
quantity of air is torn off (or sucked from) the bubble corner and becomes
stretched all around the wake. The result, commonly called the {\it skirt},
is a sheet of air aligned with the interface between the wake and the
external medium, which remains attached to the bubble (e.g.,
\cite{hnatbuck76}). A similar feature can be identified in our numerical
simulations coinciding with the streamers or tongues of vorticity of both
signs mentioned in the last section. In contrast to the air bubbles, the {\it
skirt} here is made of magnetized material.  During the whole rise the
magnetic field in the skirt is continuously compressed and stressed both by
the wake and external flows.
\clearpage\onecolumn
\thispagestyle{empty}
\begin{figure}
\vskip \textheight
%\centerline{\psfig{figure=figure14.eps,height=1.0\textheight}}
\caption{Tube and wake structure for four cases with different
  initial pitch angle but otherwise equal initial conditions. From left to
  right, the values of $\Psimax$ were $13.9^\deg$, $7^\deg$ (i.e., the case
  studied in the rest of the paper), $2.5^\deg$ and $0^\deg$. The lower
  panels show the general flow field and the contours of magnetic intensity
  for $\Blong$. The upper panels represent the velocity relative to the apex
  of the tube and the corresponding equipartition line (as defined in
  \Sec4.3.2).  The transition from a convex tube back to a flat and then to a
  concave one for decreasing initial $\Psimax$ (i.e., as the Weber number
  increases) is apparent. In the untwisted case, no {\it head} is left but,
  rather, a bridge joining the two vortex rolls of the wake.
\label{fig_comparison_initial_pitch}}
\end{figure}
\clearpage\twocolumn\noindent

\subsubsection{The tail}
\label{sec_tail}

All along the central symmetry plane, the rising tube is leaving behind a
{\it tail} of weakly magnetized matter (Fig.~\ref{fig_rise}). Its upper part
is compressed by the two rolls of the wake. In fact, the {\it tail} is not
only magnetized, but also has non-zero vorticity. This can be seen both in
the distribution of Fig.~\ref{fig_vort_angle_12} (maximum of positive
vorticity at the right end of the figure) and, as a light shadow, in the
vorticity map of Fig.~\ref{fig_vorticity_halo}. Similar structures are seen
behind moving pair of vortices as well as in bubble experiments
(e.g. \cite{wegenerparlange73}). The total amount of magnetic flux which is
lost by the tube to the tail is small (only 6 percent of the total flux).

\subsubsection{The secondary rolls at the edge of the
tube}\label{sec_secondary_rolls}  

Close to the lateral edges of the tube (see Fig.~\ref{fig_second_roll} and
the last panel in Fig.~\ref{fig_vorticity_halo}) small secondary rolls are
built from time to time; they grow for a while and then they are released to
the downflow.  These features are reminiscent of the secondary rolls observed
near the detachment point of boundary layer around solid cylinders
(\cite{taphuoc80}; \cite{taphuocbouard85}).  These secondary rolls strongly
perturb the skirt and the wake when they are released: this is the main cause
for the irregular shape of the wake in the last panel of Figs.~\ref{fig_rise}
and \ref{fig_vorticity_halo}.

%%%%%%%%%%%%%%%%%%%%%%%%%%%%%%%%%%%%%%%%%%%%%%%%%%%%%%%%%%%%%%%%%%%%%%
%%%%%%%%%%%%%%%%%%%%%%%%%%%%%%%%%%%%%%%%%%%%%%%%%%%%%%%%%%%%%%%%%%%%%%
\section{The rise of tubes with different initial degrees of twist}
\label{sec_split}
\begin{figure}[t]
\plotone{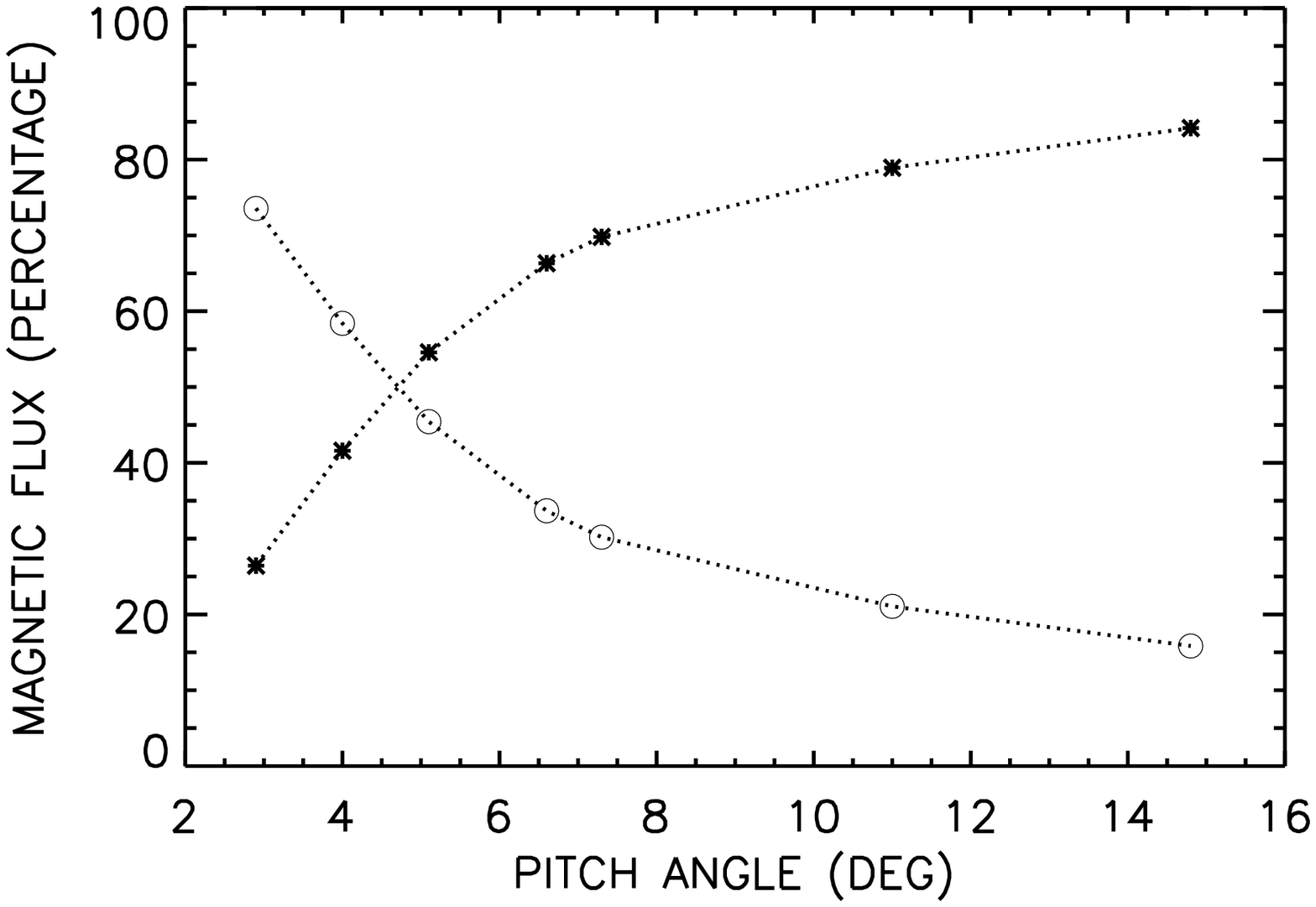}
\caption{Percentage of the initial magnetic flux remaining in the
  head of the tube once the terminal velocity has been reached (stars) as a
  function of $\Psimax$. Shown is also the percentage of flux that has been
  dragged to the wake (circles). \label{fig_remaining_flux}}
\end{figure}

The initial value of $\Psimax$ for the tube whose rise we have described in
the foregoing chapters was close to the value given by the criterion
(\ref{eq_min_pitch}), namely $\Psimax = 6^\deg$. This criterion provides a
lower bound for the twist necessary to withstand the destruction of the tube
through vorticity generation in its interior, external pressure forces, etc;
we do not know yet how sharp that bound is. In the present section we
investigate the transition from the low-twist to the high-twist regimes. We
conclude that the change from one to the other is gradual and takes place
over a range of, say, $5$ degrees in $\Psimax$, around the value given by the
right hand side of Eq.~\ref{eq_min_pitch}. In
Fig.~\ref{fig_comparison_initial_pitch} we show the relative flow, $\vrelvec$
(upper panels) and the contours of longitudinal magnetic intensity, $\Blong$
(lower panels), for four cases with $\Psimax = 13.9^\deg\;, 7^\deg\;,
2.5^\deg$ and $0^\deg$, respectively. There is a clear gradation of
properties and evolutionary patterns along the four columns of the
figure. The amount of flux remaining in the {\it head} 
of the tube is an
increasing function of $\Psimax$. This can be seen in
Fig.~\ref{fig_remaining_flux}: the stars, representing the percentage of the
initial flux which remain in the head, go from more than $80\%$ for $\Psimax
= 13.9^\deg$ to about $20\%$ for $\Psimax=2.5^\deg$ and (trivially) $0\%$ for
$\Psimax = 0^\deg$.  In this figure, the change between the two regimes
occurs between, say, $3^\deg$ and $8^\deg$.

Not only the size, but also the shape of the remaining head varies markedly
from one case to the other (Fig.~\ref{fig_comparison_initial_pitch}, upper
panels). For $\Psimax = 13.9^\deg$, the tube maintains its roundish shape all
along. This is specially striking at the rear of the tube: the rising return
flow between the two rolls in the wake cannot dent the back of the tube. This
case is closest to the motion of a rigid cylinder in a fluid. For $\Psimax$
decreasing down to $2.6^\deg$, the back of the tube becomes first flat and
then concave through the action of the return flow.  This transition
coincides with the increase of the magnetic Weber number up across unity:
$We$ is 0.19, 0.74 and 5.75 for $\Psimax = 13.9^\deg$, $7^\deg$ and
$2.5^\deg$, respectively.  In the untwisted case, there is no longer a neat
separation between tube interior and external flows. The rolls in the wake
now contain 70\% of the original magnetic flux, with most of the rest
contained in the upper arch linking them.  The evolutionary pattern in this
case is similar in many respects to the results of Longcope et al. (1996).

Further insight into the structural differences for the tubes of
Fig.~\ref{fig_comparison_initial_pitch} can be obtained by comparing
\begin{figure}[H]
\
\vskip 0.6\textheight
%\plotone{figure16.eps}
\caption{Initial and asymptotic profiles of the distribution of
  magnetic energy in the transverse field, $\etran$, along the vertical axis
  of symmetry compared, in both cases, with the asymptotic kinetic energy
  density $\ekinas$ at the apex of the tube. The asymptotic values were
  calculated at $t=5$. The different curves correspond to cases with initial
  pitch angle $\Psimax=13.9^\deg$ (dashed), $\Psimax=7^\deg$ (solid) and
  $\Psimax=2.5^\deg$ (dotted).
\label{fig_energy}}
\end{figure}
\noindent the energy density of the transverse field, $\etran$ with the
kinetic energy 
reached asymptotically by the head of the tube. The order of magnitude of the
latter is $(\Blong^2/8\pi) \, \tilder \, (1+\tan^2\Psi)$.  This varies little
(about $6\%$ only) between $\Psimax = 2.5^\deg$ and $\Psimax = 13.9^\deg$.
The magnetic energy of the transverse field, in contrast, varies by a large
factor (approx.~$30$) between those two extremes.  For the comparison
(Fig.~\ref{fig_energy}), we plot now the distribution of $\etran$ along the
vertical axis at times $t=0$ (upper panel) and $t=5$ (lower panel), divided
in both cases by the kinetic energy density at the upper stagnation point of
the tube in the asymptotic regime, $\ekinas$ (calculated in the figure at
time $t=5$). In the case with the highest $\Psimax$ (dashed line), at $t=0$
only a thin skin at the outermost tube boundary is below the terminal kinetic
energy level; we expect a skin of roughly that size to be dragged to the
wake. For $\Psimax=7^\deg$ (solid line) and $\Psimax=2.5^\deg$ (dotted line),
most (or all) of the tube at time $t=0$ has $\etran$ below the $\ekinas$
level. However, in the initial phases of the evolution the transverse field
is substantially intensified through the compression and stretching phenomena
explained in \Sec\ref{sec_rise}. This allows the {\it head} of the tube to be
formed and its energy $\etran$ to be at (or above) the $\ekinas$ level in the
asymptotic regime (Fig.~\ref{fig_energy}, dotted line in the second plot).
%
%%%%%%%%%%%%%%%%%%%%%%%%%%%%%%%%%%%%%%%%%%%%%%%%%%%%%%%%%%%%%%%%%%%%%%
%%%%%%%%%%%%%%%%%%%%%%%%%%%%%%%%%%%%%%%%%%%%%%%%%%%%%%%%%%%%%%%%%%%%%%

%%%%%%%%%%%%%%%%%%%%%%%%%%%%%%%%%%%%%%%%%%%%%%%%%%%%%%%%%%%%%%%%%%%%%%
%%%%%%%%%%%%%%%%%%%%%%%%%%%%%%%%%%%%%%%%%%%%%%%%%%%%%%%%%%%%%%%%%%%%%%
\section{Discussion}
\label{sec_discussion}

The results presented in this paper can be discussed from a twofold
perspective. First, they can be applied to the theory of the magnetic
activity in the Sun, trying to understand different aspects of the rise of
magnetic flux across the convection zone. On the other hand, the simulations
described here bear a strong resemblance to the results of laboratory and
numerical experiments on the motion of air bubbles and rigid cylinders; it
would be as well to clarify similarities and differences. The following
subsections discuss different aspects of those topics.

\subsection{The parameters and distribution of basic physical
variables}\label{sec_adeq_param_distr}

The numerical calculation of the present paper is necessarily idealized.  The
calculations were done in a high-beta regime with small ratio of radius to
pressure scaleheight. In thus far, they are within the parameter regime of
expected for the magnetic tubes in the deep convection zone. The latter have
a larger $\beta$, by a factor between $10$ and $100$, than adopted here but
many qualitative features of the rise should be similar in both cases. The
only diffusive process considered, though, the ohmic resistivity, together
with the numerical viscosity and diffusion, yield laminar flows in and around
the tube with Reynolds number of a few hundred. This cannot be expected to
hold in the actual Sun: there, the rise of the tube will be accompanied by
turbulent flows (with, in particular, a turbulent wake).

In spite of the foregoing, a conclusion seems unavoidable: for the transport
of the magnetic flux to the surface in the form of buoyant magnetic flux
tubes to be effective, the latter must be twisted from the early stages of
their rise. We have shown for a tube with a density deficit of order the full
isothermal value ($\Delta\rho/\rho \sim -1/\beta$) that an average pitch
angle around $\sin^{-1}[(\rtube/\hp)^{1/2}]$ is indeed necessary for the tube
to 
withstand the various deforming agents.  Otherwise, they lose most of their
magnetic flux to the trailing wake. As shown by earlier authors, the vortex
rolls of the wake can easily end up moving horizontally rather than
rising. The condition on the minimum pitch angle may be less stringent,
though, if we consider that the rise may actually be driven by the Parker
instability in tubes which are stored in a neutrally buoyant equilibrium
(\cite{caligarietal95}). The instability occurs with the upgoing mass
elements being driven by a vertical force which remains well below
$\rho\,g/\beta$ while the amplitude is not too large.  To a driving force of
amount $f\,\rho\,g/\beta$ ($f \ll 1$) corresponds through (\ref{eq_min_pitch})
a minimum pitch angle which is about a factor $f^{1/2}$ smaller than for
isothermal straight tubes. For example, in the case discussed in the paper of
\cite{caligarietal95}, $f < 0.04$ for the top of the rising loop in the
lowermost $10,000$ km of the rise, i.e., for distances equivalent to about
$5$ times the tube radius.

Even if the tubes are strongly buoyant [\thinspace$|\Delta \rho/\rho| =
\hbox{O}(1/\beta)$\thinspace] from the beginning, their time evolution may
depend to some extent on the precise distribution of the density deficit in
the tube at time $t=0$. A gaussian profile, as used in this paper, yields a
fast 
deformation of the initial magnetic configuration, followed by internal
torsional oscillations. This is directly associated with the torque of the
buoyancy force throughout the tube interior.  A top-hat profile ($\rho_i <
\rho_e$ but $\Delta\rho=$const), on the other hand, has no buoyancy torques
associated except at the tube boundary. Yet, the time evolution
of such a tube for smooth pitch angle distributions like
(\ref{eq_pitchassympt}) or (\ref{eq_pitchgauss}) is qualitatively
similar to the cases studied in this paper, as has been tested through 
a series of numerical experiments. This can be understood in different
ways: first, the hydrodynamic forces associated with the external flow depend
on the speed of rise of the tube, which is a function of the average buoyancy
rather than of the precise shape of the $\Delta\rho$ profile. To counteract
them, one needs to have a sufficient pitch angle in the tube interior. On the
other hand, the gravitational term in the vorticity equation (\ref{eq_vort})
for a top-hat profile is concentrated at the periphery of the $\Delta\rho$
distribution, but is correspondingly more intense.  To counteract it, it is
necessary to have a large current density $\jlong$, which can only be
achieved if $\Btran$ is sufficiently intense, at least close to the tube
boundary.  Through both arguments one arrives at criteria basically like
Eq.~(\ref{eq_min_pitch}). 

\subsection{Twisted tubes in the convection zone: three-dimensional
effects}\label{sec_disc_conv_zone} 

If the magnetic tubes that yield active regions at the surface must be
twisted already in the early stages of the rise, the Sun should have a
mechanism to routinely produce twisted tubes in the dynamo layers. If one
relaxes the condition of symmetry along the tube imposed in the present
paper, one may think that the twist could result through torsional shearing.
This might come about as a natural subproduct of the generation of vorticity
explained in this paper, if it took place at different rates on the different
cross sections of the tube along the axis.  This can be the case, for
instance, if the originally horizontal tube develops an omega-loop shape in
which a stretch of it rises while the rest remains at the original level. If
the tube was originally untwisted (or only weakly twisted), the rising
sections would tend to turn into vortex tube pairs which contain most of the
original magnetic flux. Now, in contrast to the 2D case, this rotation
produces a transverse field component in the flanks of the rising section of
the loop. If enough $\Btran$ is built up, the rotation in the vortices could
be braked. Yet, a simple calculation shows that for this mechanism to be at
all effective, the footpoint separation, $\lambda$, of the omega loop has to
be small, in fact smaller than one pressure scaleheight. More precisely, if
we require that the tube develops a transverse component of the level
required by the criterion (\ref{eq_min_pitch}) after rising a distance
equivalent to a few of its own radii, we obtain
\begin{equation}\label{eq_max_loop_foot_separation}
\frac{\lambda}{\hp} \lesssim \left(\frac{\rtube}{\hp}\right)^{1/2}\;.
\end{equation}
This is a very small footpoint separation: the ratio of the Lorentz force
associated with the curvature of the field lines to the buoyancy force in
such a tube is at least of order $\hp/\lambda$. Consequently, it is not easy
to {\it raise} a loop that is narrow enough for a sufficient pitch angle to
develop at all.  This is an indication that the tubes possibly have to be
formed with the necessary amount of twist before they start to rise. 

An interesting possibility for generating magnetic tubes with a non-zero
total twist has been discussed by \cite{cattaneoetal90}: they consider a
Rayleigh-Taylor unstable slab of plasma with horizontal magnetic field such
that the angle subtended by the field vector with a fixed horizontal axis
monotonically changes with depth.  Upon development of the instability, the
layer yields magnetic tubes with non-zero degree of twist. In their
experiment, the authors showed how the resulting tubes were more resistant to
deformation through the surrounding flows than in the corresponding untwisted
case (\cite{cattaneohughes88}).  A corresponding three-dimensional problem
has been calculated by \cite{matthewsetal95}. They show how an unstable layer
with a parallel horizontal field produces magnetic tubes with non-zero
vorticity. Through nonlinear interaction, these tubes arch as they rise in a
vertical plane, thereby becoming twisted.

Another open question concerns the fate of a twisted tube as it rises across
successive density scaleheights in the convection zone. The internal and
external densities are essentially equal for most of the rise (except for the
small relative difference of order up to $1/\beta \ll 1$): the tube thus
expands and its field intensity weakens by several orders of magnitude along
the journey (see \cite{morenoinsertis86}, 1992, 1997b). In fact, the
beginning of this process can be seen by comparing the maximum field
intensity of the three panels of Fig.~\ref{fig_pitch_angle}. The rate of
decrease associated with an off-axis expansion of the tube, though, is
different for the longitudinal and transverse components of the field
(\cite{parker79}, \Sec9). A rough but simple argument based on the
conservation of flux in 2D yields a rate of change of the twist following an
approximate law $\tan\Psi \propto \rtube$. If so, the tubes could reach the
upper convection zone with very large degrees of twist, possibly such that
they can become kink unstable. Yet, three-dimensional effects could render
that simple law of little use. The stretching of the tube apex in the
longitudinal direction, for instance, a phenomenon common in Parker-unstable
rising loops, may reduce the level of twist there. Also, the conservation of
the total helicity in the tube could put an upper limit to the number of
turns of the field lines around the axis at any single place. However it is,
large pitch angles should be a common appearance in rising tubes when they
reach photospheric levels. The sheared field structures observed in emerging
active regions (\cite{litesetal95}, \cite{lekaetal96}) may be a consequence
of this (see also \cite{tanaka91}, \cite{kurokawa89} and \cite{rustkumar96}).

\subsection{Magnetic tubes, air bubbles and rigid cylinders: similarities and
differences}
Along this paper we have compared our results to the laboratory experiments of
flow past rigid cylinders and air bubbles at $Re \cong 200$ -- $300$. In the
following we summarize the similarities and differences found. 

{\it Magnetic tubes and air bubbles.\/} Remarkable similarities are: (a) the
general shape and structural features (skirt, central tail) (b) the {\it
protection} of the interior of the rising object through surface
tension. These similarities are all the more striking given the difference in
buoyancy [density deficit $|\Delta\rho/\rho| \ll 1$ for the tubes,
$\hbox{O}(1)$ for the bubbles] and in the physical source of the surface
tension (capillary effects vs. jump in the tangential field component). As a
result, defining the Weber number on the basis of the corresponding surface
tension mechanism, one can formulate a common law of dependence of structural
properties on $We$.

{\it Magnetic tubes and rigid cylinders.\/} A clear similarity concerns the
boundary layer: both objects have a no-slip condition along the boundary.
Consequently, the relative jump of the tangential velocity across the
boundary layer is large, $[v_{tg}^{}]/v_{tg}^{} \sim \hbox{O}(1)$.
%$[{\vrel{}}_{tg}^{}]/{\vrel{}}_{tg}^{} \sim \hbox{O}(1)$. 
In contrast, there is a zero-tangential stress condition along
the fluid/air interface of a bubble; correspondingly,
$[v_{tg}^{}]/v_{tg}^{} \sim \hbox{O}(Re^{-1/2})$. As a result of the no-slip
%$[\vrel{}_{tg}^{}]/\vrel{}_{tg}^{} \sim \hbox{O}(Re^{-1/2})$. As a result,
condition, both buoyant magnetic tubes and rigid cylinders produce (and shed
downstream) secondary rolls near the point of separation of the boundary
layer. 

{\it Features specific to the buoyant magnetic tubes with twist.}  In the
magnetic tubes of the present simulations, the wake is formed in the initial
phases of the rise out of material bodily transported from the initial tube
Thus, the wake is magnetized and maintains a magnetic connection to the head
of the rising tube along time.  The kind of behavior followed by the magnetic
rope depends directly on the amount of magnetic flux incorporated into the
wake, which, in turn, depends on the initial twist. In the case of an
initially highly twisted tube ($\Psimax=13.9^\deg$), the magnetic rope is
almost rigid and its shape and the flow around it strongly remind those of a
{\it solid cylinder}. If the initial pitch angle is closer to the threshold
(\ref{eq_min_pitch}), then the tube deforms and adopt a {\it bubble}-like
shape (though it still satisfies a no-slip condition along its
boundary). Finally, if $\Psimax$ is very small, the tube behaves like a
rising {\it thermal} (Longcope et al. 1996).

%%%%%%%%%%%%%%%%%%%%%%%%%%%%%%%%%%%%%%%%%%%%%%%%%%%%%%%%%%%%%%%%%%%%%%
%%%%%%%%%%%%%%%%%%%%%%%%%%%%%%%%%%%%%%%%%%%%%%%%%%%%%%%%%%%%%%%%%%%%%%
\acknowledgments This work was partially funded through the DGES project
no.~95-0028-C of the Spanish Ministry of Education and Culture.  The
numerical calculations were carried out using the computing resources of the
Instituto de Astrof\'{\i}sica de Canarias and of the Centre de Computaci\'o i
Comunicacions de Catalunya. The authors are grateful to M. Kaisig for
providing the numerical code and for his subsequent help with it. Thanks are
also due to a large number of scientists in HAO, JILA and CORA in Boulder
(Colorado) as well as in the University of Chicago for interesting
discussions on the topic of this paper.

%%%%%%%%%%%%%%%%%%%%%%%%%%%%%%%%%%%%%%%%%%%%%%%%%%%%%%%%%%%%%%%%%%%%%%
%%%%%%%%%%%%%%%%%%%%%%%%%%%%%%%%%%%%%%%%%%%%%%%%%%%%%%%%%%%%%%%%%%%%%%

%%%%%%%%%%%%%%%%%%%%%%%%%%%%%%%%%%%%%%%%%%%%%%%%%%%%%%%%%%%%%%%%%%%%%%
%%%%%%%%%%%%%%%%%%%%%%%%%%%%%%%%%%%%%%%%%%%%%%%%%%%%%%%%%%%%%%%%%%%%%%

\end{document}